\documentclass[a4paper, 12pt]{article}  
\usepackage{pdfpages}
\usepackage{lmodern}
\usepackage{graphicx} 
\usepackage{needspace}
\usepackage{wrapfig} 
\usepackage{textcomp} 
\usepackage[utf8]{inputenc}
\DeclareSymbolFont{letters}{OML}{ztmcm}{m}{it}
\usepackage{color}
\usepackage{multicol}
\usepackage{multirow}
\usepackage[boxed]{algorithm2e}

\usepackage{rotating}
\usepackage{listings}
\usepackage{adjustbox}
\usepackage[boxed]{algorithm2e}
\usepackage{subcaption}
\usepackage{multirow} 
\usepackage{booktabs} 
\usepackage[hang,flushmargin]{footmisc}
\usepackage{pdflscape}
\usepackage{breakurl}
\usepackage[doi=only,eprint = false ,isbn=false,uniquelist=false, firstinits=true,maxcitenames=2, maxbibnames = 10000,authordate,backend=biber,natbib]{biblatex-chicago}

\DeclareFieldFormat{doi}{\href{https://doi.org/#1}{\nolinkurl{doi:#1}}}
\DeclareFieldFormat{pages}{\hspace*{-3.4pt}#1}

\DeclareNameAlias{sortname}{family-given}
\AtEveryBibitem{\clearfield{month}}
\usepackage[T1]{fontenc} 
\usepackage{geometry}
\usepackage{amsmath}
\usepackage{etoolbox} 
\usepackage{amsthm}
\usepackage{amsfonts}
\usepackage{amssymb}
\usepackage[section]{placeins}
\usepackage{diagbox} 
\usepackage{hyperref}
\usepackage{graphicx} 
\usepackage{fixmath} 
\usepackage[scaled=.90]{helvet}
\usepackage{wrapfig}
\usepackage{tikz}
        \usepackage{setspace}
\usepackage{setspace}
\makeatletter
\renewcommand\@biblabel[1]{\textbf{#1.}} 
\usepackage{enumitem}

\usepackage{pifont}

\setlist{nolistsep,leftmargin=*}
\renewcommand{\maketitle}{ 
\begin{center}
{\LARGE\@title} 

\vspace{0pt} 

{\large\@author} 
\\\@date 

\vspace{40pt} 
\end{center}
}
\begin{document}
	\singlespace
\begin{center}
	\textbf{The Role of Governmental Weapons Procurements in Forecasting Monthly Fatalities in Intrastate Conflicts: A Semiparametric Hierarchical Hurdle Model} \\
	Cornelius Fritz$^\dagger$, Marius Mehrl$^\ddagger$, Paul W. Thurner$^\ast$, Göran Kauermann$^\dagger$\hspace{.2cm}\\
	Department of Statistics, LMU Munich$^\dagger$\hspace{.2cm}\\
	Department of Government, University of Essex$^\ddagger$\hspace{.2cm}\\  Geschwister Scholl Institute of Political Science, LMU Munich$^\ast$\hspace{.2cm}
\end{center}

\begin{abstract}
	Accurate and interpretable forecasting models predicting spatially and temporally fine-grained changes in the numbers of intrastate conflict casualties are of crucial importance for policymakers and international non-governmental organisations (NGOs). Using a count data approach, we propose a hierarchical hurdle regression model to address the corresponding prediction challenge at the monthly PRIO-grid level.
	More precisely, we model the intensity of local armed conflict at a specific point in time as a three-stage process. Stages one and two of our approach estimate whether we will observe any casualties at the country- and grid-cell-level, respectively, while stage three applies a regression model for truncated data to predict the number of such fatalities conditional upon the previous two stages. Within this modelling framework, we focus on the role of governmental arms imports as a processual factor allowing governments to intensify or deter from fighting. We further argue that a grid cell's geographic remoteness is bound to moderate the effects of these military buildups. Out-of-sample predictions corroborate the effectiveness of our parsimonious and theory-driven model, which enables full transparency combined with accuracy in the forecasting process.  
\end{abstract}
	{ \textit{Keywords: Conflict Forecasting, Conflict Intensity, Forecasting, Hurdle Regression, Semiparametric Regression}}   
\section*{Introduction}

Within sub-Saharan Africa alone, the violent deaths of close to 8,500 people are attributed to state-based armed conflict in 2019 \citep{Pettersson_Öberg_2020}.  Reliable forecasts of conflict intensifications before they occur allow policymakers to take precautionary and de-escalatory measures, thus decreasing the human toll of organised violence. We present an approach that extends recent advances in the forecasting of armed conflicts (e.g., \citealp{Blair_Sambanis_2020,Chiba_Gleditsch_2017,Hegre2019, Ward_Metternich_Dorff_Gallop_Hollenbach_Schultz_Weschle_2013}) towards forecasting conflict intensity, as indicated by the number of casualties resulting from state-based conflicts. 

We follow this endeavour at the geographically highly disaggregated level of the PRIO grid cell, which is a standardised structure introduced by \citet{Tollefsen} comprising quadratic grid cells that cover the entire world at an approximate resolution of 55 $\times$ 55km. By doing so, we extend the current literature in two main regards. First, we propose to forecast the intensity of state-based fighting by employing a novel hierarchical hurdle regression model that we develop in accordance with our theoretic considerations. Second, we expand the suite of commonly used covariates to include the role of governmental weapons procurements on the country level. 

Our model rests on the idea of hurdle models \citep{Cragg1971} but advances the corresponding model class in two main aspects: we incorporate three nested stages and adopt thresholding \citep{Sheng2006}, a technique from cost-sensitive classification, to hurdle models. In this stage-wise regression, the first stage indicates whether we will observe any fatalities at the country level (country violence incidence). The second stage controls whether conditional on having any casualties at the country level, we will also observe a non-zero number of deaths in a given grid cell (cell violence incidence). Given that this is the case, the third stage then uses truncated regression to estimate the number of fatalities (cell violence intensity). 
To make predictions that are consistent with the training observations, we introduce two cutoff parameters that serve as thresholds known from cost-sensitive classifications in the first two binary stages of our model \citep{Hernandez-Orallo2012}. We set these parameters according to a metric that ensures well-calibrated predictions, i.e., predicting approximately the cumulative fatalities we observe.  



In applying our three-staged approach to forecasting local conflict intensity, we heed recent calls for more theory-based conflict prediction \citep{Blair_Sambanis_2020,Chiba_Gleditsch_2017,Cederman_Weidmann_2017} and limit the covariates in our model to a parsimonious set of theoretically motivated variables. In line with the suggestion to ``focus on the processes that produce violence closer to the moment of onset'' \citep[3]{Chiba_Gleditsch_2017}, we emphasise governmental major conventional weapons imports as a driver of conflict intensity \citep{Mehrl_Thurner_2020}. Notably, such arms transfers also increase the risk of violence to occur \citep{Pamp2018a, Magesan_Swee_2018}. However, as we discuss below, their conflict-inducing effect will not be homogeneous across all locations within a country. In particular, we expect that governmental imports of major conventional weapons increase fighting intensity but that this effect decreases with a location's distance from the capital. This is the case as remote localities are harder to reach for armies that employ major conventional weapons due to their lower ability to traverse rough terrain, increased reliance on road infrastructure, and more challenging logistics. 

In support of our modelling approach, the out-of-sample evaluation indicates that our hierarchical hurdle regression model outperforms a competitive random forest benchmark model. Furthermore, our test results suggest that the inclusion of governmental arms imports increases the predictive accuracy of the forecasts. We will henceforth use the abbreviation PGM and CM when referring to monthly observations at the PRIO-grid and country level. The corresponding spatial units are shortened to PG (PRIO-grid cell) and C (country). 

This article's remainder is structured as follows: the next section motivates the hierarchical structure and use of a hurdle model in the application case. Building on this theoretical foundation, we then formally introduce the semiparametric hierarchical hurdle model. Subsequently, we apply this model to predict local conflict intensity at the PRIO grid level. This section also presents the specification of a parsimonious suite of covariates, out-of-sample evaluation results, as well as forecasts until March 2021. The paper concludes with a discussion of possible future directions.


\section*{Theoretical Motivation}

A principal issue in modelling armed conflict is that conflict events are empirically rare. This is the case for all commonly used units of observation (state-dyads, country-years, etc.) but in particular at more granular levels of spatiotemporal resolution such as PGs. For instance, even throughout the Liberian civil war, most locations in Liberia did not experience any fighting as violence instead clustered in a few areas \citep{Hegre_Østby_Raleigh_2009}. And while resulting in over 22,000 civilian casualties, much of the violence in the Bosnian civil war was concentrated in a few infamous events, most prominently the massacres in Srebrenica and Prijedor. In contrast, other periods passed without any reported killings of civilians \citep{Schneider_Bussmann_Ruhe_2012}. The main concern here is that there is an excess of zero-observations, i.e. observations where no conflict occurs \citep{Bagozzi_2015, Beger_Dorff_Ward_2016}. Most obviously, some country-pairs, states, or subnational locations may be as good as immune to armed conflict because of their particular wealth, institutional features, or geographical attributes. For instance, it is highly improbable that Djibouti and Lesotho would ever engage in a mutual dispute or that Liechtenstein would experience civil war. But even in countries with an ongoing conflict, fighting is usually geographically confined and unlikely to reach locations such as the capital \citep{Hegre_Østby_Raleigh_2009, Buhaug_2010, Tollefsen_Buhaug_2015}. Furthermore, such excess zeroes can also occur temporally and, if left unaccounted for, threaten both inference and our ability to forecast fighting \citep{Bagozzi_2015}.  


When predicting the monthly conflict intensity at the spatially highly disaggregated PRIO grid level, this discussion has relevant implications. To begin, we can expect that as numerous countries will be at peace in either a month under observation or even across the entirety of the period we study, none of the grid cells contained within them will experience any fighting. And even when violence does occur on the country-month level, the majority of its grid cells will nonetheless see no combat. 
A descriptive analysis of the available data, covering violence at the cell-month level across Africa during the period 1990-2019, supports these expectations. First, most fatalities occur in a small subset of countries that hardly changes over time.\footnote{To be precise, almost 70$\%$ of fighting casualties occur in only four countries, namely Eritrea, Ethiopia, Sudan, and Somalia.} This implies that hierarchical structure, i.e., in which country each cell is situated, carries vital information for the prediction task. Second, the vast majority of grid cells - even within countries experiencing combat -  ($99.2\%$) are reporting zero cases. As there are more than 10,000 grid cells defined in Africa for each month, the datasets might therefore include up to 3,8 million observations. This, in turn, can posit an obstacle when estimating flexible and realistic models in a context where fast as well as precise and interpretable predictions are needed.

In predicting PGM-level fighting, we focus on the external procurement of major conventional weapons as a critical factor that allows governments to engage in and escalate armed conflict. Existing studies identify these imports as drivers of conflict onset \citep{Pamp2018a, Magesan_Swee_2018} but also emphasise their potential to intensify current fighting as they increase governmental forces' ability to pin down the enemy and engage in decisive battles \citep{Caverley_Sechser_2017, Mehrl_Thurner_2020}. Hence, the procurement of weapons is a driver of both the occurrence and intensity of conflict. That being said, governmental arms imports are a country-level factor while we are interested in predicting grid-level fighting; this is particularly relevant as grid cells differ in terms of their potential to experience conflict or be exposed to government-owned heavy weaponry. Namely, the state's reach  - and hence rebels opportunity to challenge it - varies over its territory, with locations far away from the capital being the most difficult to govern and thus most suitable for rebellion \citep{Tollefsen_Buhaug_2015, Boulding_1962, Buhaug_2010}. Such remoteness may, in turn, mainly affect the power projection ability of forces employing major conventional weapons given their comparatively lower ability to traverse rough terrain, higher reliance on road infrastructure, and more challenging logistics. In addition to increasing the country-level occurrence and intensity of fighting, governmental arms imports may hence also determine combat severity at the more local level. We thus use the external procurement of weapons to identify which countries experience fighting and, in interaction term with a PG's distance from the capital, to predict the local occurrence and intensity of violence within these countries.   

\section*{Hierarchical Hurdle Regression}
\label{sec:model}

Stemming from this discussion, we propose a forecasting model, which can incorporate the hierarchical data structure, given by each PG allocation in a country, as well as appropriately deal with the high rate of excess zeros. These aims are mirrored in two model characteristics: a stage-wise formulation and an application of cutoff values for prediction. Our model explicitly assumes that state-based PG fatalities occur only in countries that we predict to have at least one fatality. We make this prediction based on a binary regression model on the country level and introduce a cutoff value to obtain binary predictions. Given that this prediction forecasts fatalities on the country level, we progress to a second binary decision on the PG level to determine if we will observe at least one case in the respective cell. Similar to the first binary choice, we use an additional cutoff value to attain binary predictions. Provided that this binary result is again positive at the PGM level, we capture the realised count by a truncated distribution defined over the positive natural numbers.\footnote{As we do not use the standard 25 battle death threshold for armed conflict, the first two classification stages of our model should not be interpreted as identifying conflict and non-conflict units.}

\subsection*{Model Formulation}

For a precise notation, we order each grid cell according to the country it is situated in and, therefore, define $y_{ijt}$ to be the observed number of state-based fatalities in PRIO-grid $i$ situated in country $j$ and month $t$, with $i = 1, ..., n_j$, $j = 1, ..., n$ and \newline $t \in \mathcal{T} = \lbrace \text{January }1990, ...,\text{March }2021 \rbrace$. In our application we set $n = 55$ (number of countries) and $n_j$ denotes the number of cells located in country $j = 1, ..., n$. Since our model combines both the country and grid level, let $y_{\cdot jt} = \sum_{i = 1}^{n_j} y_{ijt}$ be the corresponding observation aggregated at the country level.  In accordance with the abbreviations introduced earlier, we write CM $jt$ and PGM $ijt$ to shorten the corresponding observations. Further, we define a binarised version of $y_{\cdot jt}$ by $\tilde{y}_{\cdot jt}$, hence $\tilde{y}_{\cdot jt} = y_{\cdot jt} >0$. Within this notation, the aim of the prediction task is to forecast $\Delta_{ijt}^s$ defined by: 
\begin{align}
    \log(y_{ijt} +1) - \log(y_{ijt-s} +1) \text{ with } (t,s) \in \lbrace (\text{October } 2020, 2), ... , (\text{March } 2021, 7) \rbrace, \label{eq:delta}
\end{align}
with data given until $t-s$. Since the sole stochastic component of \eqref{eq:delta} is $\log(y_{ijt} +1)$ it suffices to model the observed counts at the point in time $t$. We tackle this endeavour with six models that only differ by the assumed lag structure of $s$ months between the measurement of all covariates and the target variable (with $s = 2, ..., 7$). Since we have data until August 2020, the one-step-ahead forecasts of these models yield the predictions needed in \eqref{eq:delta}. Simply put, the prediction of the time-step $s$ months into the future translates into a lag of $s$ months between covariates and the target variable. Without a loss of generality, we hence formulate our model for the arbitrary delay structure of $s$ months.

%

In our three-stage hurdle regression model, we decompose the probability distribution of the random variable $Y_{ijt}$ into two binary decisions and a truncated counting model. In stage 1 we start with a binary classification model on the country level in which the target variable is the random variable $Y_{\cdot jt}$, indicating whether we observed at least one fatality in CM $jt$.  Conditional on having observed at least one fatality in CM $jt$, the consecutive two stages encompass a standard hurdle regression model \citep{Mullahy1986}. Therefore, stage 2 constitutes another binary decision determining if we observe at least one fatality in PGM $ijt$, while stage 3 models the count of deaths conditional on having observed at least one death in the respective cell. For stage 3, we utilise a truncated counting distribution. Mathematically, the resultant probability model of this stage-wise approach can be stated as the joint bivariate probability of $Y_{ijt}$ and $\tilde{Y}_{\cdot jt}$ with delay structure $s$ by: 
\begin{align}
    \mathbb{P}( Y_{ijt} = y_{ijt}, \tilde{Y}_{\cdot jt} = \tilde{y}_{ \cdot jt} \mid x_{\cdot jt-s}^{(1)}, x_{ijt-s}^{(2)}, x_{ijt-s}^{(3)}) = \begin{cases}
    \big(1- \pi_{\cdot jt}^{(1)}\big)& \tilde{y}_{\cdot jt}= 0, y_{ijt} = 0 \\
    \pi_{\cdot jt}^{(1)} \big(1- \pi_{ijt}^{(2)}\big)& \tilde{y}_{ \cdot jt} = 1, y_{ijt} = 0 \\
    \pi_{\cdot jt}^{(1)} \pi_{ijt}^{(2)} f_{tr}^{(3)}(y_{ijt})& \tilde{y}_{\cdot jt}= 1, y_{ijt} > 0 \\
    0 & \text{else}
    \end{cases}, 
    \label{eq:model}
\end{align}
where $ \pi_{\cdot jt}^{(1)}$ is the probability of observing at least one fatality in CM $jt$,  $\pi_{ijt}^{(2)}$ the  probability of observing at least one fatality in PGM $ijt$, and $f_{tr}^{(3)}(y_{ijt}) = \frac{f_{C} (y_{ijt})}{1 - f_{C}(0)}$ the density of a zero-truncated version of a discrete random variable with density $f_{C}(y_{ijt})$. The bivariate nature of \eqref{eq:model} is a technical necessity and due to $\tilde{Y}_{\cdot jt}$ being the target variable in the first stage. However, when using model \eqref{eq:model} to obtain forecasts we only consider the prediction of $y_{ijt}$ and view the prediction of $\tilde{y}_{ \cdot jt} $ as a byproduct. The required quantities are obtained from the three sub-models, which each incorporate covariates that are measured at time point $t-s$ and denoted by $ x_{\cdot jt-s}^{(1)}, x_{ijt-s}^{(2)},$and $x_{ijt-s}^{(3)}$ for the commensurate sub-models. The corresponding dependency is implicitly assumed to guarantee a less cluttered notation. The marginal expected value of $Y_{ijt}$ defined in \eqref{eq:model} is given by 
\begin{align}
    \mathbb{E}(Y_{ijt}) = \frac{\pi_{\cdot jt}^{(1)} \pi_{ijt}^{(2)}}{1-f_{C}(0)} \mathbb{E}_C(Y_{ijt}), \label{eq:mean}
\end{align}
where $\mathbb{E}_C(Y_{ijt})$ is the expected value of the counting variable with density $f_{C}(y_{ijt})$ and $Y_{ijt}$ is defined in \eqref{eq:model}.\footnote{This result follows from the direct calculation of the marginal density of $Y_{ijt}$ and the application of the total law of probability.} 

\subsection*{Stage-wise Specification}

In general, the quantities defining \eqref{eq:model}, namely $ \pi_{\cdot jt}^{(1)}, \pi_{ijt}^{(2)}$ and $f_{tr}^{(3)}(y_{ijt})$, can be specified separately by arbitrary regression techniques. Our specification aims to be as flexible as needed while, at the same time, providing transparent and interpretable forecasts and coefficients. Therefore, we suggest the usage of generalised additive mixed models \citep{Ruppert2003a, ruppert2009,wood2017}, which entail the following distributional assumptions: 
\begin{enumerate}
    \item The two binary targets in stage 1 and 2,  $y_{\cdot jt}> 0$ and $y_{ijt} > 0$, follow a Binomial distribution with the corresponding success probabilities $\pi_{\cdot jt}^{(1)}$ and $\pi_{ijt}^{(2)}$.
    \item The truncated counting variable in stage 3 follows a truncated Poisson disctribution with (untruncated) mean $\lambda_{ijt}^{(3)}$.
\end{enumerate}

We parameterise the means of the three corresponding distributions in terms of stage-specific lagged covariates, $ x_{\cdot jt-s}^{(1)}, x_{ijt-s}^{(2)}, x_{ijt-s}^{(3)}$. As our model is of a semiparametric nature, we incorporate these covariates in each stage as having either a linear ($L$) or nonlinear ($NL$) effect. Accordingly, we decompose all covariates along their effect type, e.g., in the third stage $x_{ijt-s}^{(3)} = \big( x_{ijt-s}^{(3,L)}, x_{ijt-s}^{(3,NL)}\big)$ for covariates with linear and non-linear effects (the same holds for stage 1 and 2). The sum of all effects results in stage-wise linear predictions, which in our specification are given by: 
\begin{align}
    \eta_{\cdot jt}^{(1)}  &=  \big(\theta^{(1,L)}\big)^\top x_{\cdot jt-s}^{(1,L)} + \sum_{\tilde{x} \in x_{\cdot jt-s}^{(1,NL)}} f(\tilde{x}) + u_j \nonumber\\
     \eta_{ijt}^{(2)} & =  \big(\theta^{(2,L)}\big)^\top x_{ijt-s}^{(2,L)} +  \sum_{\tilde{x} \in x_{ijt-s}^{(2,NL)}} f(\tilde{x})\label{eq:np}\\ 
     \eta_{ijt}^{(3)}  &=  \big(\theta^{(3,L)}\big)^\top x_{ijt-s}^{(3,L)} + \sum_{\tilde{x} \in x_{ijt-s}^{(3,NL)}} f(\tilde{x}),  \nonumber
\end{align}
where $\theta^{(k,L)}$ with $k = 1, 2, 3$ are linear coefficients to be estimated, $f(\cdot)$ smooth non-linear functions specified through basis functions, e.g., P-Splines \citep{eilers1996}, and $u_j$ is a Gaussian country-specific random effect.\footnote{For our application the definition of those sets of covariates ($x_{ijt-s}^{(k,L)}, x_{ijt-s}^{(k,NL)}$ with $k = 1,2,3$) as well as the specification of smooth components are given in Appendix A. } Let the stage-specific parameters defining all components in \eqref{eq:np} be $\theta^{(1)}, \theta^{(2)}, \theta^{(3)}$.  

Besides, we accommodate possible restrictions on the means, i.e.,$\pi_{\cdot jt}^{(1)} \in [0,1]$ and $\lambda_{ijt}^{(3)} \in \mathbb{R}^+$, by transforming the linear predictors defined in \eqref{eq:np} by a response function \citep{Wedderburn1972}. While we apply the inverse logit transformation for the binary regressions,  the exponential function is used for the truncated Poisson model. The relations between the stage-wise means and linear predictors are therefore given by: 
\begin{align*}
    \pi_{\cdot jt}^{(1)} = \frac{\exp \big\lbrace \eta_{\cdot jt}^{(1)} \big\rbrace} {1- \exp \big\lbrace \eta_{\cdot jt}^{(1)} \big\rbrace}, \pi_{ijt}^{(2)} = \frac{\exp \big\lbrace \eta_{ijt}^{(2)} \big\rbrace} {1- \exp \big\lbrace \eta_{ijt}^{(2)} \big\rbrace}, \text{ and } \lambda_{ijt}^{(3)} = \exp \big\lbrace \eta_{ijt}^{(3)} \big\rbrace.
\end{align*}
 
Under conditional independence between the model stages, we estimate $\theta^{(1)}, \theta^{(2)}, \theta^{(3)}$ through three separate generalised additive mixed models. The contribution of $y_{ijt}$  to the joint likelihood of time-step $s$ is: 
\begin{align}
    \mathcal{L}(\theta^{(1)}, \theta^{(2)}, \theta^{(3)} \mid y_{ijt},y_{\cdot jt}) = &f_{Bin}^{(1)}\big(y_{\cdot jt}> 0 \mid \theta^{(1)}\big)^{1/n_j} f_{Bin}^{(2)}\big(y_{ijt}> 0 \mid \theta^{(2)}\big)^{\mathbb{I}(y_{\cdot jt}> 0)} \nonumber\\ &f_{trP}^{(3)}\big(y_{ijt} \mid \theta^{(3)}\big) ^{\mathbb{I}(y_{ijt}> 0)}, 
    \label{eq:like}
\end{align}
where $f_{Bin}(y   \mid \theta)$ is the density of a Bernoulli random variable and  $f_{trP} (y\mid \theta)$ the density of a zero-truncated Poisson distribution both parametrised as described in the distributional assumptions of the previous paragraph. The two levels of analysis at the CM and PGM level entail the inclusion of a scale factor $1/n_j$. Heuristically this is necessitated by the fact that from any $y_{ijt} >0$ it follows that $y_{\cdot jt}>0$ must hold, hence all other observations in the respective countries $y_{\tilde ijt} ~ \forall ~  \tilde i \in \lbrace 1, ..., n_j \rbrace$ and $\tilde i \neq i$ do not include any additional information regarding the first stage.  The product of \eqref{eq:like} over all observations in the training set yields a complete likelihood. To achieve the best trade off between complexity and simplicity, we penalise the roughness of all nonlinear terms and include side constraints to ensure identifiability as detailed in \citet{Wood2020}. For details on the implementation, we refer to Annex D. 

\subsection*{Sparse Predictions by Thresholding}

To obtain sparse predictions from the estimated model, we introduce two additional cutoff parameters. In classification tasks, cutoff values are commonly used to transform the probability output of a model, here the probability of a unit to observe at least one fatality, into a binary classification whether we predict at least one fatality for a country or not \citep{Domingos1999MetaCostAG}. For instance, assuming that this value is 0.4, we would predict at least one death in a country if the respective predicted probability is above 0.4. Theoretically, the threshold should be 0.5, yet for applications with strongly imbalanced data such as ours or cost-sensitive misclassification, e.g. the diagnosis of cancer, \citet{Sheng2006} proposed to learn this value as an additional tuning parameter. 

Along these lines, we extend thresholding to our multi-stage hurdle regression (the application to standard hurdle regression follows naturally) and introduce two additional parameters that serve as thresholds for the first two binary stages of our model. Building on the notion of crossing hurdles and the corresponding model class's name, we call these parameters hurdles and denote them by $\tau_1$ and $\tau_2$. By applying those hurdles, we only predict non-zero values on the PGM level if the corresponding country probability from the first stage is higher than $\tau_1$ and the probability to have at least one case in the PG from the second stage is above $\tau_2$. Having specified $\pi_{\cdot jt}^{(1)},\pi_{ijt}^{(2)}$, and $f_{tr}^{(3)}(y_{ijt})$ by generalised additive mixed models, we obtain predicted values for each stage, denoted by $\hat\pi_{\cdot jt}^{(1)},\hat\pi_{ijt}^{(2)}$, and $ \hat\lambda_{ijt}^{(3)}$. With these values, the prediction $\hat{y}_{ijt}$ under $\tau_1$ and $\tau_2$ is given by: 
\begin{align}
   \hat{y}_{ijt}(\tau_1,\tau_2)  =
   \begin{cases}
   0 &(\hat\pi_{ijt}^{(1)}< \tau_1) \lor (\hat\pi_{ijt}^{(2)} < \tau_2)\\
   \hat\lambda_{ijt}^{(3)} & \text{else}
   \label{eq:pred_th}
\end{cases}
\end{align}


 Generally, there are numerous ways to set these thresholds, and the decision highly depends on the desired forecast. For instance, if the MSE of the predictions should be as low as possible, the thresholds can be picked so that the misclassification rates of the two corresponding binary models are minimised. Since this approach often leads to overly sparse predictions, we seek well-calibrated predictions in our application. In other words, we want to predict approximately the amount of fatalities that are observed. We thus opt to minimise the following loss in terms of $\tau_1$ and $\tau_2$: 
\begin{align}
   loss(\tau_1, \tau_2 \mid y) = |\log(\hat{y}(\tau_1, \tau_2) +1)^\top \mathbf{1} -     \log(y+1)^\top  \mathbf{1}|, \label{eq:loss}
\end{align}
where $\hat{y}(\tau_1, \tau_2)$ defines the vector stacking all predictions obtained through applying \eqref{eq:pred_th}, $y$ the stacked observed counts, and $\mathbf{1}$ a vector of ones. To minimise \eqref{eq:loss} in a time-efficient fashion, we apply an algorithm for global optimisation, namely differential evolution \citep{Das2011}. Overall, these hurdles make our modelling framework more flexible and adaptable to specific goals that can be set by the practitioner. 

\subsection*{Data Partitioning}

\begin{table}[t!]
	\centering
	\begin{tabular}{l | c  c}
		& Evaluation & Forecast \\ \hline
		Pre-Training & Jan. 1990  to $(t - s - 1)$ & Jan. 1990 to Jul. 2020 \\
		Calibration & $t - s$ &  Aug. 2020 \\
		Training & Jan. 1990  to $t - s$ &  Jan. 1990  to Aug. 2020 \\
		Test & $t$ &  Oct. 2020 to Mar. 2021 \\
	\end{tabular}
	\caption{Periodisation of data for expanding window evaluation at time point\newline $t \in \lbrace \text{Jan. 2017}, ..., \text{Dec. 2019} \rbrace$ and $s \in \lbrace 2, ..., 7 \rbrace$ and forecasting.}
	\label{tab:my_label}
\end{table}

\begin{algorithm}[t!]
	\SetAlgoLined
	\KwResult{Predictions at $\mathcal{T} = \lbrace \text{January } 2017, ...,\text{December }2019 \rbrace$}
	\emph{Let} $\mathtt{data}$ be the complete dataset including all information until December 2019 \\
	\For{$t \in \mathcal{T}$}{
		\For{$s \in \lbrace 2, ..., 7\rbrace$}{
			\textbf{1. Data Preparation} 
			\begin{enumerate}[label=(\alph*)]
				\item  $\mathtt{data\_tmp} =  \mathtt{data}[date \leq t]$\newline  Delete data that is unknown and unnecessary for the present forecast
				\item  $\mathtt{data\_tmp} = \mathtt{lag\_cov}(\mathtt{data\_tmp}, s)$ \newline Induce lag structure of $s$ months
				\item  $ (\mathtt{pre\_train,calibrate,test}) = \mathtt{split}(\mathtt{data\_tmp})$\newline Split data as shown in Table \ref{tab:my_label}
			\end{enumerate}
			\textbf{2. Pre-Estimation} (Train models with pre-training data up to $t-s-1$)
	   		\begin{enumerate}[label=(\alph*)]
	   			\item  $\mathtt{model\_1} = \mathtt{estimate (pre\_train, stage = 1)}$
	   			\item  $\mathtt{model\_2} = \mathtt{estimate (pre\_train, stage = 2)}$
	   			\item  $\mathtt{model\_3} = \mathtt{estimate (pre\_train, stage = 3)}$ 
	   		\end{enumerate}
   			\textbf{3. Calibration} (Calibrate the thresholds $\tau_1$ and $\tau_2$ with data from $t-s$)
   			\begin{enumerate}[label=(\alph*)]
   				\item  $\mathtt{pred\_calib} = \mathtt{pred} (\mathtt{model\_1}, \mathtt{model\_2}, \mathtt{model\_3, data = calib})$ \newline Prediction of $\hat{\pi}_{\cdot j(t-s)}^{(1)}, \hat{\pi}_{ij(t-s)}^{(2)}$, $\hat{\lambda}_{ij(t-s)}^{(3)}$ for observations in  $t-s$
   				\item $\mathtt{thresholds = DEoptim(pred\_calib, loss = cal\_loss)}$\newline $\mathtt{cal\_loss}$ is given by \eqref{eq:loss} and $\mathtt{DEoptim}$ a minimisation routine \newline based on a genetic algorithm 
   			\end{enumerate}
   			\textbf{4. Estimation}  (Train models with training data up to $t-s$)
   			\begin{enumerate}[label=(\alph*)]
   				\item  $\mathtt{train = join(pre\_train, calibrate)}$
   				\item  $\mathtt{model\_1} = \mathtt{estimate (train, stage = 1)}$
   				\item  $\mathtt{model\_2} = \mathtt{estimate (train, stage = 2)}$
   				\item  $\mathtt{model\_3} = \mathtt{estimate (train, stage = 3)}$
   			\end{enumerate}
   			\textbf{5. Prediction}  (Generate and save forecasts for $t$) 
   			\begin{enumerate}[label=(\alph*)]
   				\item   $\mathtt{pred\_per\_stage} = \mathtt{pred} (\mathtt{model\_1}, \mathtt{model\_2}, \mathtt{model\_3, data = test})$\newline Prediction of $\hat{\pi}_{\cdot jt}^{(1)}, \hat{\pi}_{ijt}^{(2)}$, $\hat{\lambda}_{ijt}^{(3)}$ for observations in  $t$
   				\item  $\mathtt{pred\_final = apply\_thresholds(pred\_per\_stage,thresholds)}$ \newline  $\mathtt{apply\_thresholds}$ is given by \eqref{eq:pred_th} 
   				\item  Save $\mathtt{pred\_final}$
   			\end{enumerate}
		}
	}
	\caption{Pseudo-Code of the evaluation forecasts.}
	\label{evaluation_forecasts}
\end{algorithm}

To provide the forecasts in \eqref{eq:delta} with data available until August 2020, we employ an one-step-ahead procedure on the models with $s = 2, ..., 7$. For instance, we forecast the counts in November 2020 by lagging the covariates by $s = 3$ months as we are given data until August 2020. In the same manner, we calculate expanding-window evaluation forecasts for January 2017 to December 2019 for all time-steps ($s = 2, ..., 7$) to measure the out-of-sample performance of our model. This procedure 
adequately replicates the real forecasting situation \citep{Richardson2020}. In accordance with the real forecasts in \eqref{eq:delta}, the prediction task of the evaluation is thus given by:  
\begin{align}
    \log(y_{ijt} +1) - \log(y_{ijt-s} +1) ~  \forall  ~ s = 2, ..., 7, i = 1, ..., n_j, j = 1, ..., n \text{ and } t  \in \mathcal{T}_{Evaluation}, \label{eq:delta_eval}
\end{align}
where $\mathcal{T}_{Evaluation} = \lbrace \text{January } 2017, ...,\text{December } 2020\rbrace$ and given data until $t-s$. The entire procedure for the latter type of prediction is summarised in Algorithm \ref{evaluation_forecasts}.\footnote{To clarify, data until a specific point in time $t$ describes observations where the target variable was measured at $t$ and the covariates at $t-s$. As we provide expanding-window forecasts, we run through the inner loop of Algorithm \ref{evaluation_forecasts} (sub-task 1 through 5) for the prediction of each month ($\mathcal{T} = \lbrace \text{January } 2017, ...,\text{December }2019 \rbrace$) and time-step ($s = \lbrace 2, ..., 7 \rbrace$), where we set the periodisation according to Table \ref{tab:my_label}. } 

For predicting the fatalities at time point $t$ and time-step $s$, we use data until $t -s$ which we split into pre-training and calibration dataset according to Table \ref{tab:my_label} (sub-task 1). We start by estimating the model with the data from the pre-training data  (sub-task 2). Consecutively, we optimise \eqref{eq:loss} given the calibration data  (sub-task 3) and re-estimate the model on the training data  (sub-task 4). Finally, we transform the forecasted fatalities at $t$ by applying \eqref{eq:delta_eval} to $\Delta_{ijt}^s$ and save the results (sub-task 5).

\section*{Application}
\label{sec:app}

We next employ the hierarchical hurdle regression model formulated above to forecast monthly changes in the intensity of fighting across the African continent. In particular, we focus on governmental arms imports and, on the PG level, their interaction with a location's distance from the capital as theoretically critical covariates. We first discuss the construction of these and other covariates. Consecutively, we present out-of-sample evaluations of our proposed model as well as forecasts for October 2020 until March 2021.      

\subsection*{Description of Covariates}

The  covariates, denoted by $x_{\cdot jt-s}^{(1)}, x_{ijt-s}^{(2)}$ and $x_{ijt-s}^{(3)}$, can be specified for each stage individually. 
As discussed above, each regression model we calculate has a specific delay between target and regressor $s \in \lbrace 2, ..., 7 \rbrace$. For clarity, we define all covariates here before applying these model-specific lags.  


For our key independent variable, governmental imports of major conventional weapons, we use data from the SIPRI Arms Transfer Database \citep{sipridata2017}, covering global arms transfers from 1950 to the present. We construct two yearly variables from this dataset as we distinguish between ``short-term'' and ``long-term'' imports of weapons. The former cover the weapon imports during the present and previous year, while the latter sums up the procurements between three and ten years before the present year. We denote the corresponding yearly covariates by $x_{\cdot j y}^{LT}$ and $x_{\cdot j y}^{ST}$ for country $j$ and year $y$ to define them as: 
\begin{align*}
    x_{\cdot jy}^{ST} &= \log \big( x_{j,y}^{\text{MCW Import}} + x_{j,y-1}^{\text{MCW Import}}  \big) \\
     x_{\cdot jy}^{LT} &= \log \big(\sum_{\tilde{y} = y - 10}^{y-2} x_{j,\tilde{y}}^{\text{MCW Import}} \big), 
\end{align*}
where $x_{j,y}^{\text{MCW Import}} $ is given by the yearly import of major conventional weapons measured in TIVs \citep{siprimeth}. The temporal scale of the resultant covariates is then transformed from yearly to monthly by setting $x_{\cdot jt}^{ST} = x_{\cdot jy}^{ST}$ and $x_{\cdot jt}^{LT} = x_{\cdot jy}^{LT}$ for all months $t$ within year $y$. In other words, short-term imports of MCW reflect the total strategic value of the weapons procured from abroad in the two years preceding an observation while long-term imports indicate the aggregate value of arms obtained in the eight years before that. As discussed above, we include these variables in all three stages of the model, but in stages 2 and 3 interact them with the location's distance to the capital \citep{Weidmann}, which accordingly also enters stages 2 and 3 as a covariate. To capture the belligerents' (potential) structural military power \citep{Mehrl_Thurner_2020}, we further include governmental military expenditures in all three stages \citep{SIPRI}.

Additionally, our model contains three further groups of selected covariates. First, we account for spatial dynamics in our data by including a country-specific Gaussian random effect in the first stage as well as a smooth spatial effect of a location's average longitude and latitude in all three stages. Second, we include a smooth time trend, dummy variables for month effects, the time since the last fatality as well as the total number of fatalities in the previous month resulting from organised violence to take temporal dynamics into account. Third, we incorporate a small number of covariates that have been shown to be key structural predictors of armed conflict onset and intensity (see \citealp{Hegre2013, Hegre2019}). 
In Appendix A, a complete list of all included covariates is given with the respective data sources and transformations.

\subsection*{Results}

\subsubsection*{Out-Of-Sample Evaluation}

	\begin{figure}[t!]
		\centering
		\includegraphics[width=0.9\linewidth]{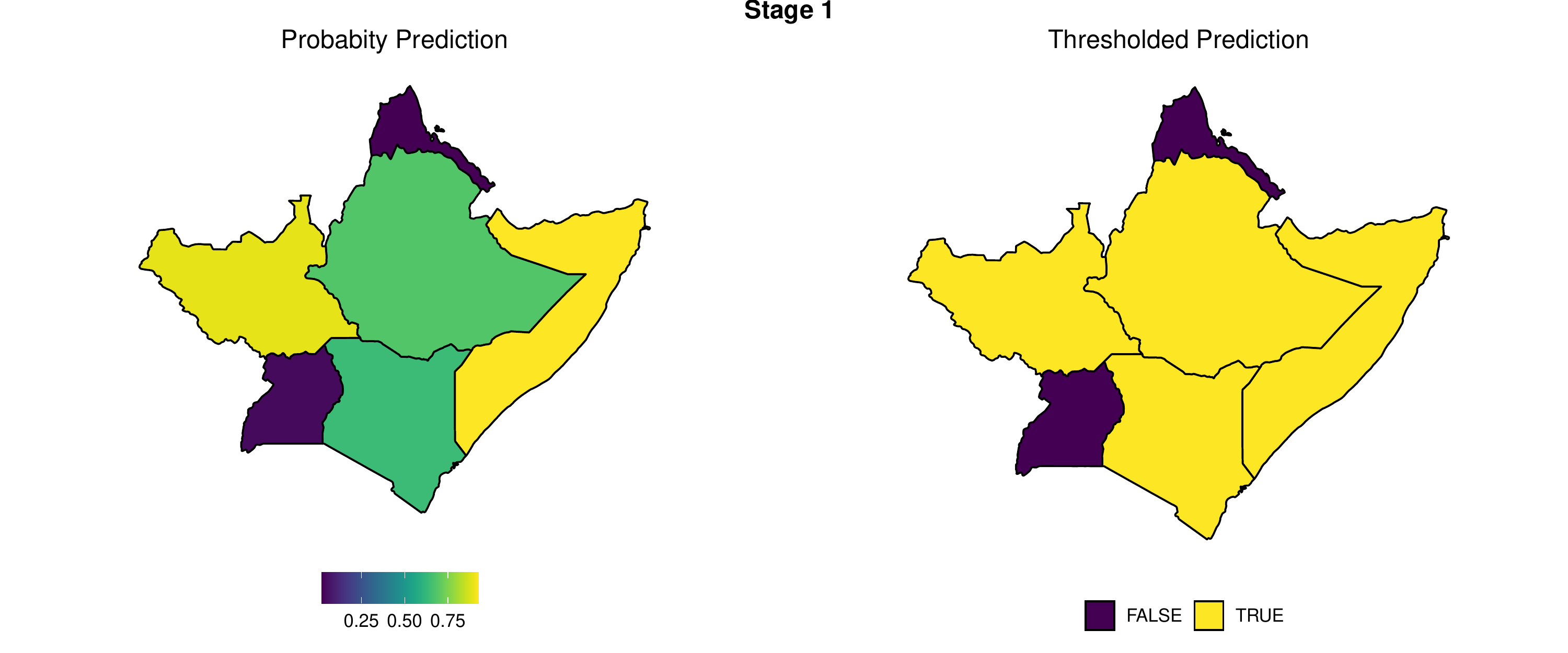}
		\includegraphics[width=0.9\linewidth]{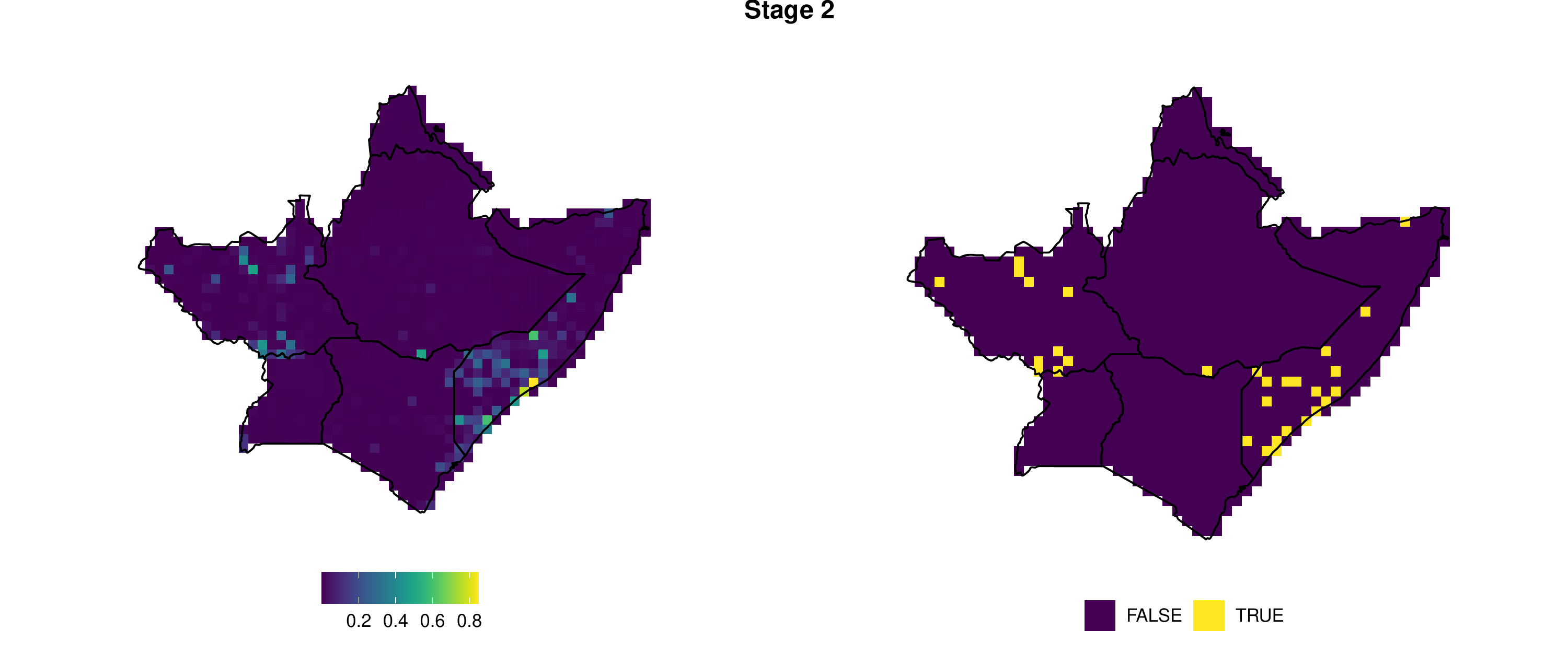}
		\includegraphics[width=0.85\linewidth]{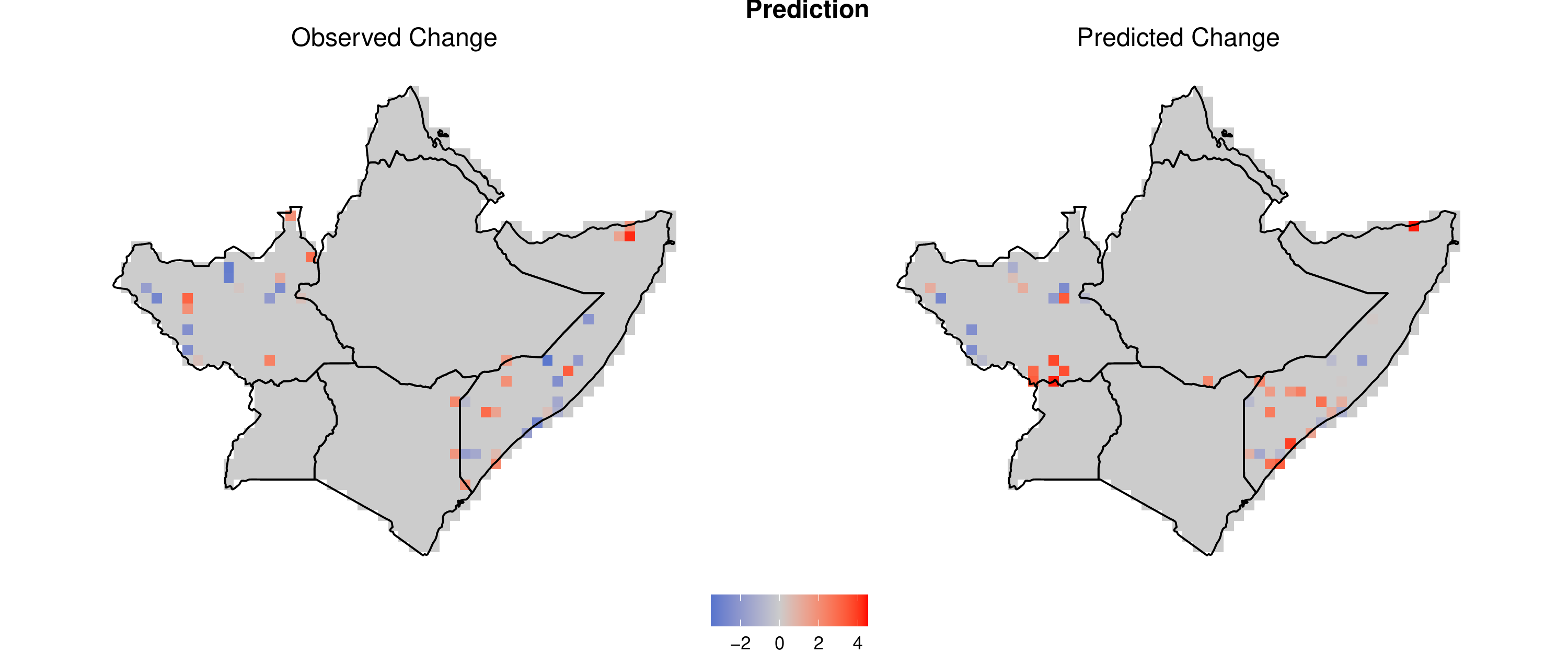}
		\caption{Observed and predicted changes in fatalities $\Delta_{ijt}^s$ with $s = 2$ in June 2018 in a sub-sample of countries. The predictions are separated along Stage 1 (first row), Stage 2 (second row), and final predicted change to April 2018 (third row). }
		\label{fig:prediction}
	\end{figure}
	
All three steps needed to obtain predictions for the number of state-based casualties in June 2018 with data until April 2018 ($s = 2$) are illustrated in Figure \ref{fig:prediction} for a sub-sample of five countries in North-Eastern Africa (Eritrea, Ethiopia, Uganda, Somalia, Kenya, South Sudan). The first row depicts the predicted probability of country fighting incidence as well as whether the model expects violence to occur after applying the calibrated threshold. The second row presents these quantities for the cell level and shows that in this example, the optimised threshold probabilities at which the model expects fighting to take place differ from 0.5 ($\tau_1 =0.563,\tau_2 =  0.263$). In the third row, the final out-of-sample predicted fatalities in June 2018 are transformed to changes in conflict intensity. Here, we show the predictions on the right and the observed changes on the left side. A comparison of these two final maps suggests that our model is generally successful in identifying which countries will experience armed conflict as well as in predicting where those fatalities will occur. Notably, the model correctly identifies Eritrea and Uganda as not experiencing conflict in this month. On the sub-national scale, we predict both the location and the direction of, for instance, changes in the intensity of fighting in western South Sudan and southern Somalia rather well. At the same time, the maps demonstrate that our model was unable to forecast what seems like relatively isolated flare-ups of violence in, e.g. eastern Kenya. 

	\begin{table}[t!]
    \centering
    \renewcommand{\arraystretch}{1}
\begin{tabular}{l|ccc|ccc}
 & \multicolumn{3}{c|}{MSE} &  \multicolumn{3}{c}{TADDA} \\
s & MCW & No MCW & Benchmark & MCW & No MCW & Benchmark\\
\hline
2 & \textbf{0.034} & 0.034 & 0.045 & \textbf{0.016} & 0.016 & 0.151\\
3 & \textbf{0.035} & 0.037 & 0.046 & \textbf{0.016} & 0.017 & 0.138\\
4 & \textbf{0.036} & 0.040 & 0.050 & \textbf{0.017} & 0.017 & 0.140\\
5 & \textbf{0.038} & 0.042 & 0.048 & \textbf{0.017} & 0.018 & 0.151\\
6 & \textbf{0.037} & 0.043 & 0.050 & \textbf{0.017} & 0.018 & 0.142\\
7 & \textbf{0.038} & 0.048 & 0.052 & \textbf{0.017} & 0.020 & 0.151\\
\end{tabular}
    \caption{Out-of-sample MSE and TADDA scores from the hierarchical hurdle model with and without the MCW-related covariates as well as the benchmark model.}
    \label{tab:my_second_table}
\end{table}


To evaluate the model in a more principled manner, we compare the out-of-sample performance of 1) the fully specified hierarchical hurdle regression model, 2) the same model without MCW imports and military capacities, and 3) a competitive benchmark model based on a random forest. Let $\hat{y}_{ts}$ be the stacked vector of all predictions at time point $t$ with data until $t-s$, while the stacked predicted changes from \eqref{eq:delta_eval} are denoted as $\Delta \hat{y}_{ts}$ and the matching observed values are $y_{t}$ and $\Delta y_{ts}$. Under this notation, the MSE and TADDA scores for $s = 2, ..., 7$ are given by: 
 \begin{align*}
     MSE_s &= \frac{1}{ \# \mathcal{T}_{Evaluation}} \sum_{t \in \mathcal{T}_{Evaluation}}\frac{1}{ \# y_t} (\hat{y}_{ts} - y_t)^2\\
     TADDA_s^\epsilon &=  \frac{1}{\# \mathcal{T}_{Evaluation}} \sum_{t \in \mathcal{T}_{Evaluation}} \frac{| \Delta {y}_{ts} - \Delta \hat{y}_{ts}| + |\Delta \hat{y}_{ts}|\mathbb{I} \big(\Delta \hat{y}_{ts}^{(\pm)}  \neq \Delta {y}_{ts}^{(\pm)}\big) \mathbb{I}( |\Delta \hat{y}_{ts} - \Delta {y}_{ts} | > \epsilon) }{\#\Delta {y}_{ts} },
 \end{align*}
 where $\mathbb{I}(\cdot)$ is the indicator function, $y^{(\pm)}$ denotes the sign of all values in vector $y$ and $\# y$ the dimension of $y$, i.e., $\# y = n$ for $y \in \mathbb{R}^n$. Finally, $\epsilon$ is a value set at 0.048 to roughly correspond to a change of $5\%$ in state-based fatalities. The resulting values are presented in Table \ref{tab:my_second_table}. Overall, the MSE and TADDA scores are consistently higher for the benchmark model than for both hierarchical hurdle models. This indicates that accounting for the hierarchical structure of and excess zeroes within our data improves the ability to forecast conflict escalation and de-escalation.  Simultaneously, these results suggest that weapons transfers and capabilities are key covariates when predicting the dynamics of state-based fighting. The out-of-sample validation hence provides clear support for our proposed model in terms of utilised model class and model specification.


\subsubsection*{Forecasts}

	\begin{figure}[t!]
		\centering
		\includegraphics[width=0.7\linewidth]{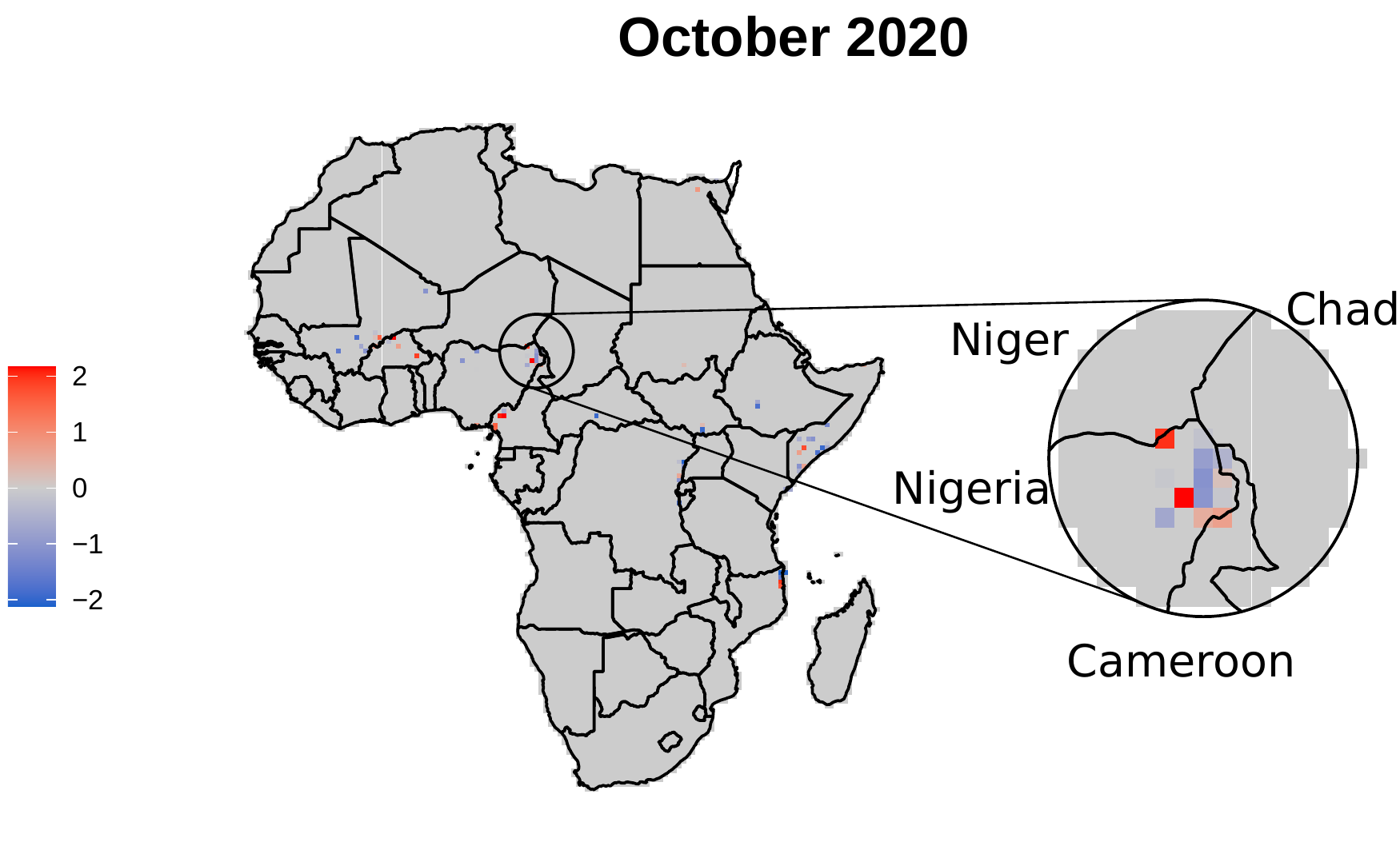}
		\includegraphics[width=0.7\linewidth]{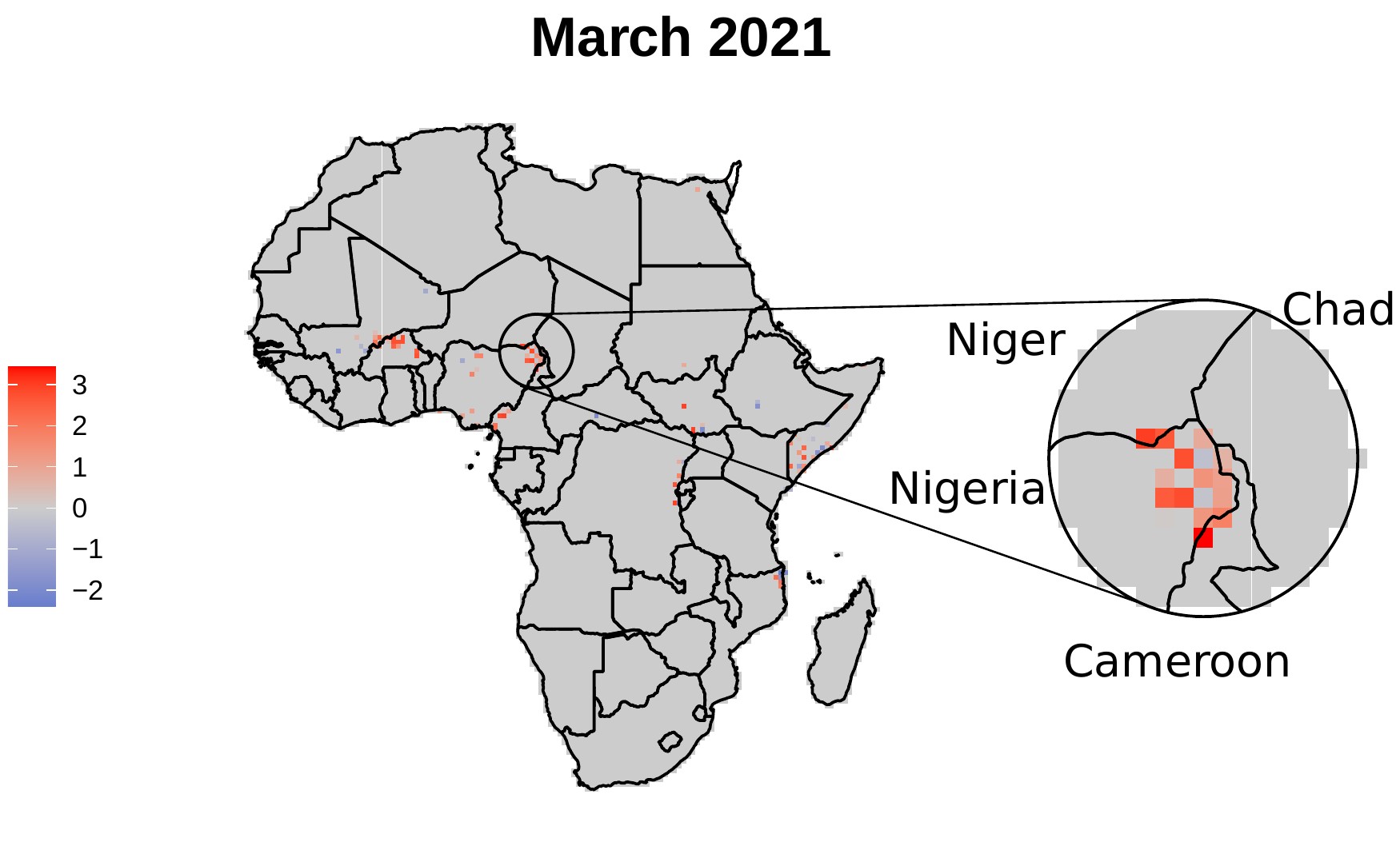}
		\caption{Forecasted log changes $\Delta_{ijt}^s$ in fatalities to October 2020 ($s = 2$) and March 2021 ($s = 7$), focused on the border region between Nigeria, Niger, Cameroon, and Chad.}
		\label{fig:forecast}
	\end{figure}
As already explained, we can exploit the proposed model to make monthly predictions up to March 2021. 
Figure \ref{fig:forecast} graphically presents the starting and end points of these forecasts. Similar to Figure \ref{fig:prediction}, changes in conflict intensity are predicted to geographically cluster in a few regions which are often more distant to the corresponding country's capital. One of these regions is the border area between Cameroon, Chad, Niger, and Nigeria, where Boko Haram has been particularly active. There, our model predicts violence to both de-escalate and escalate within the forecasting period. For October 2020, the forecast indicates that violence will decrease in most Nigerian locations but increase in PGs very close to Nigeria's borders with Cameroon and Niger. In contrast, it expects violence to increase almost across the board in March 2021 as many locations both within Nigeria and bordering Cameroon and Niger are predicted to experience a rise in casualty numbers. The forecast concurrently expects Chad to remain unaffected by these changes in violence.

While less publicised than the Boko Haram case, Mozambique has also experienced the rise of a violent Islamist insurgency since its first attacks in 2017 \citep{Morier-Genoud_2020}. These insurgents have claimed numerous attacks in the country's northeastern Cabo Delgado province and our model also predicts casualty numbers to shift in this region.\footnote{See Figure 7 in Appendix C.} It expects violence to move southwards along the coast as it predicts casualties to decrease in some of the northernmost areas of the province but to increase closer to its administrative centre Pemba. Violence is thus expected to spread further within Cabo Delgado, reaching areas where the terrorist organisation has previously shown little activity, but not across the border into Tanzania.        

\begin{table}[t]
\centering \small
\begin{tabular}{l lc ccc} 
 &Covariate & Estimate & Std. Error & t-value & p-value \\    \hline
  \multirow{3}{*}{\begin{sideways}Stage 1\end{sideways}}  
 & Military Expenditure & 0.0624 & 0.0511 & 1.2227 & 0.2215 \\ 
  &Long-Term Import of MCW & -0.0898 & 0.0452 & -1.9851 & 0.0471 \\ 
  & Short-Term Import of MCW & 0.1104 & 0.0285 & 3.8673 & 0.0001 \\ \hline
 \multirow{6}{*}{\begin{sideways}Stage 2\end{sideways}}  & Military Expenditure &  0.0923 & 0.0173 & 5.3437 & $<$ 0.0001 \\ 
 & Distance to Capital (CD)  & 0.0001 & 0.0001 & 1.2348 & 0.2169 \\ 
 & Long-Term Import of MCW (LT) & -0.0291 & 0.0229 & -1.2678 & 0.2049 \\ 
 & Short-Term Import of MCW (ST) & -0.0643 & 0.0179 & -3.5986 & 0.0003 \\ 
 & LT×CD  & -0.0002 & 0.0000 & -9.5322 & $<$ 0.0001 \\ 
 & ST×CD  & 0.0000 & 0.0000 & 1.4582 & 0.1448\\ 
   \hline
  \multirow{6}{*}{\begin{sideways}Stage 3\end{sideways}}   
  &  Military Expenditure & 0.1870 & 0.0438 & 4.2712 &  $<$ 0.0001\\ 
  & Distance to Capital (CD)  & 0.0020 & 0.0002 & 11.2872 & $<$ 0.0001 \\ 
  & Long-Term Import of MCW (LT)  & 0.6844 & 0.0519 & 13.1837 & $<$ 0.0001 \\ 
 &  Short-Term Import of MCW (ST) & -0.0140 & 0.0418 & -0.3349 & 0.7377\\ 
  & LT×CD & -0.0009 & 0.0001 & -12.7715 & $<$ 0.0001 \\ 
 & ST×CD & -0.0002 & 0.0001 & -3.5346 & 0.0004 \\ 
   \hline
\end{tabular}
\caption{The subset of the parametric estimates regarding the governmental procurement of weapons with $s = 2$ and training data from January 1990 to August 2020.} 
\label{tab.gam}
\end{table}

Finally, we present a subset of the linear estimates underlying the October 2020 forecasts in Table \ref{tab.gam} (hence $s =2$) to examine to what extent the external procurement of weapons drives these predicted changes in violence. These results support our general expectation that arms imports fuel conflict. As suggested, at least long-term arms imports increase local conflict intensity, but this relationship becomes weaker for locations farther away from the capital. This conclusion is displayed in the negatively signed and statistically significant interaction terms in stage three. While there is little evidence that recent imports of MCW affect fighting at the PG-level, the estimates of stage 1, in contrast, allow drawing the inference that such transfers can trigger the country-level occurrence of fighting while longer-term imports are associated with a lower probability of lethal violence. These results demonstrate that our model not only outperforms the benchmark model in out-of-sample predictions but additionally enables practitioners to understand how specific covariates affect the forecasts. As the findings on the varying effects of short- and long-term imports of MCW on the occurrence and intensification of fighting show, these results also have the potential to add new nuance to existing research.

\section*{Conclusion}

Predicting the escalation and de-escalation of armed conflict at fine-grained spatio-temporal levels is a crucial concern to researchers and policymakers alike. In this article, we develop and apply a hierarchical hurdle regression model that explicitly accounts for two key features of conflict event data, namely hierarchical structure and excess zeroes. Inspired by recent calls to develop theoretically motivated conflict forecasting models, we, in particular, emphasise the role of governmental weapons imports as a potential driver of both country- and local-level fighting. Out-of-sample evaluations attest that both of our methodological and substantive contributions increase the ability to predict conflict (de-)escalations. We employ our modelling approach to forecast changes in the log-transformed number of casualties for October 2020 to March 2021. Showcasing the interpretability of our model, we further examine to what extent arms imports trigger and fuel violence. As a result, we find evidence that such transfers impact the occurrence and intensification of fighting in nuanced, previously unobserved ways. Overall, our model hence not only provides practitioners and policymakers with forecasts on future conflict escalation but can furthermore inform them about the principal drivers of - and hence levers to address - this fighting.


More generally, the hierarchical hurdle approach presented here will also be of interest to conflict scholars and forecasters due to its adaptability. For instance, its hurdle regression step could easily be appended to the binary ensemble model proposed in \citet{Hegre2019} by merely adding a conditional truncated layer onto the predicted posterior probabilities. Similarly, the semi-parametric specification we utilised in all three stages can easily be exchanged for arbitrary machine learning techniques. 

Finally, this research points to the added value of including theoretically grounded processual covariates such as arms imports when forecasting conflict. As such, future work on conflict forecasting should leverage increasingly available fine-grained event data on other theoretically relevant covariates such as crop production, weather events, or migration. To better account for the serial dependencies and resulting self-exciting behaviour, another possible enhancement of the covariates would be to incorporate self-exciting terms in the flavour of \citet{Porter2010}. Further, armed conflict is not the only type of event which is essential to forecast and occurs within a hierarchical data structure and with a wealth of excess zeroes. Hence, it would be fruitful to apply and extend our model to study any hierarchically clustered count data. Examples of such data are manifold and include the occurrence and local intensification of different types of conflict (e.g., one-sided or non-state), protests or infectious disease outbreaks. 



\label{sec:conclusion}
\printbibliography
\end{document}


\singlespace
\begin{center}
	\textbf{Appendix \\ The Role of Governmental Weapons Procurements in Forecasting Monthly Fatalities in Intrastate Conflicts: A Semiparametric Hierarchical Hurdle Model} \\
	Cornelius Fritz$^\dagger$, Marius Mehrl$^\ddagger$, Paul W. Thurner$^\ast$, Göran Kauermann$^\dagger$\hspace{.2cm}\\
	Department of Statistics, LMU Munich$^\dagger$\hspace{.2cm}\\
	Department of Government, University of Essex$^\ddagger$\hspace{.2cm}\\  Geschwister Scholl Institute of Political Science, LMU Munich$^\ast$\hspace{.2cm}
\end{center}

\appendix 
\tableofcontents
\newpage

\clearpage
\FloatBarrier

\section{Covariates}
\FloatBarrier

\subsection{Incorporation and Specification}
To provide completely transparent forecasts, we here give additional information on the model specification that comprises the decomposition of the stage-specific covariates into having either linear or nonlinear effects. Consecutively, we state the specification of the incorporated smooth components. 

The decompositions introduced in formula (4) of the main article are given by: 
\begin{align*}
    x_{ijt-s}^{(1)} &= \big( x_{ijt-s}^{(1,L)}, x_{ijt-s}^{(1,NL)}\big) \\
    x_{ijt-s}^{(1,L)} &= \lbrace \text{Intercept,NS Fatality in last month, OS Fatality in last month,} \\
    &\hspace{0.74cm}\text{SB Fatality in last month, SB Fatality Count in last month, Population,} \\
    &\hspace{0.74cm}\text{GDP, Polity IV Index, Military Expenditure, Long-Term Import of MCW,} \\
    &\hspace{0.74cm}\text{Short-Term Import of MCW, Dummy variables for each month} \rbrace \\
    x_{ijt-s}^{(1,NL)}  &= \lbrace t,\text{Time since NS Fatality, Time Since OS Fatality, }  \\
   &\hspace{0.74cm} \text{ Time Since SB Fatality, Longitude/Latitude} \rbrace \\
     x_{ijt-s}^{(2)} &= \big( x_{ijt-s}^{(2,L)}, x_{ijt-s}^{(2,NL)}\big) \\
    x_{ijt-s}^{(2,L)} &= \lbrace \text{Intercept,NS Fatality in last month, OS Fatality in last month,} \\
    &\hspace{0.74cm}\text{SB Fatality in last month, SB Fatality Count in last month, Nightlights, } \\
    &\hspace{0.74cm}\text{Infant Mortality Rate, Intensity SB, Population,GDP, } \\
    &\hspace{0.74cm}\text{Polity IV Index, Military Expenditure, Distance to Capital, } \\
      &\hspace{0.74cm}\text{Long-Term Import of MCW, Short-Term Import of MCW, }  \\
        &\hspace{0.74cm}\text{Dummy variables for each month} \rbrace \\
     x_{ijt-s}^{(2,NL)}  &= \lbrace t,\text{Time since NS Fatality, Time Since OS Fatality, }  \\
   &\hspace{0.74cm} \text{ Time Since SB Fatality, Longitude/Latitude} \rbrace \\
         x_{ijt-s}^{(3)} &= \big( x_{ijt-s}^{(3,L)}, x_{ijt-s}^{(3,NL)}\big) \\
    x_{ijt-s}^{(3,L)} &= \lbrace \text{Intercept,NS Fatality in last month, OS Fatality in last month,} \\
    &\hspace{0.74cm}\text{SB Fatality in last month, Nightlights, } \\
    &\hspace{0.74cm}\text{Infant Mortality Rate, Intensity SB, Population, GDP, } \\
    &\hspace{0.74cm}\text{Polity IV Index, Military Expenditure, Distance to Capital, } \\
      &\hspace{0.74cm}\text{Long-Term Import of MCW, Short-Term Import of MCW, }  \\
        &\hspace{0.74cm}\text{Dummy variables for each month} \rbrace \\
     x_{ijt-s}^{(3,NL)}  &= \lbrace t,\text{Time since NS Fatality, Time Since OS Fatality, Time Since SB Fatality,  }  \\
   &\hspace{0.74cm} \text{Intensity SB, Longitude/Latitude} \rbrace \\
\end{align*}

Plugging these values into (4) of the main article leads to the linear predictors and full model equations. While we parametrise the temporal trend by Gaussian process smooths \citep[see][]{fahrmeir2013}, P-splines are used for all other univariate functions \citep[][]{eilers1996}. Finally, for the spatial components, bivariate tensor product of thin-plate splines \citep{wood2003} are applied that take the average longitude and latitude as the parameters (further information is given in \citealp{wood2013}). For the regression model with an arbitrary temporal delay of $s$ months, the decomposition and parametrisation of all included covariates are summarised in Table \ref{tab:my_label}. Definitions of $Y_{\cdot jt}$ and $Y_{ijt}$ are given in the main article, while SB, OS, and NS abbreviate the state-based (SB), one-sided (OS), and non-state (NS) conflict in line with \citet{Hegre2019}.

\begin{table}[t!]
    \centering
    \begin{tabular}{c|c|c| c | c}
         Variable & Incorporation & Stage 1 & Stage 2 & Stage 3 \\\hline
         $Y_{\cdot jt} > 0$ & Target & $\checkmark$ & & \\
         $Y_{ijt} > 0$ & Target &  &$\checkmark$ & \\
         $Y_{ijt}$ & Target &  &  &$\checkmark$ \\
         Temporal Trend & GP Smooth & $\checkmark$ & $\checkmark$ & $\checkmark$ \\
         Month Effect & Dummy Variables & $\checkmark$ & $\checkmark$ & $\checkmark$ \\
         Months Since last SB Fatality& P-Spline & $\checkmark$ & $\checkmark$ & $\checkmark$ \\
         Months Since last OS Fatality & P-Spline & $\checkmark$ & $\checkmark$ & $\checkmark$ \\
         Months Since last NS Fatality & P-Spline & $\checkmark$ & $\checkmark$ & $\checkmark$ \\
         SB Fatality in $t -s$ & Dummy Variable & $\checkmark$ & $\checkmark$ & $\checkmark$ \\
         OS Fatality in $t -s$ & Dummy Variable & $\checkmark$ & $\checkmark$ & $\checkmark$ \\
         NS Fatality in $t -s$ & Dummy Variable & $\checkmark$ & $\checkmark$ & $\checkmark$ \\
         SB Fatality Count in $t-s$ & Linear Effect & $\checkmark$ & $\checkmark$ &  \\
         SB Fatality Count in $t-s$ & P-Spline & &  & $\checkmark$ \\
         Spatial Effect & Tensor TP-Spline  & $\checkmark$ & $\checkmark$ & $\checkmark$ \\
         Country Effect & Random Effect  & $\checkmark$ &  &  \\
         Military Expenditure & Linear Effect  & $\checkmark$ & $\checkmark$ & $\checkmark$  \\
         Distance to Capital (CD)  & Linear Effect &  & $\checkmark$ & $\checkmark$  \\
         Short-Term Import of MCW (ST) & Linear Effect & $\checkmark$ & $\checkmark$ & $\checkmark$  \\
         Long-Term Import of MCW (LT) & Linear Effect & $\checkmark$ & $\checkmark$ & $\checkmark$  \\
         ST$\times$ CD & Interaction Effect &  & $\checkmark$ & $\checkmark$  \\
         LT $\times$ CD & Interaction Effect &  & $\checkmark$ & $\checkmark$  \\
         Polity IV Index  & Linear Effect & $\checkmark$ & $\checkmark$ & $\checkmark$  \\
         GDP  & Linear Effect & $\checkmark$ & $\checkmark$ & $\checkmark$  \\
         Population  & Linear Effect & $\checkmark$ & $\checkmark$ & $\checkmark$  \\
         Nightlights &  Linear Effect  &  & $\checkmark$ & $\checkmark$  \\
         Infant Mortality Rate &  Linear Effect  &  & $\checkmark$ & $\checkmark$  \\
    \end{tabular}
    \caption{Variables in each stage of our hierarchical hurdle regression with delay structure $s$.}
    \label{tab:my_label}
\end{table}

\FloatBarrier
\newpage
\subsection{Data Sources and Transformations}
\label{sec:covs}
\FloatBarrier

In Table \ref{tab:my_label}, we condense the sources and used transformation for all covariates incorporated in our model with delay structure $s$.  
\begin{table}[t!]
    \centering
    \begin{tabular}{c|c|c}
         Variable & Transformation & Source \\\hline
         $Y_{\cdot jt} > 0$ & Logit-Link & \citet{ucdp}\\
         $Y_{ijt} > 0$ & Logit-Link & \citet{ucdp} \\
         $Y_{ijt}$ & Log-Link &  \citet{ucdp} \\
         Temporal Trend & Identity &  - \\
         Months Since last SB Fatality & $\log(\cdot +1)$ & \citet{ucdp} \\
         Months Since last OS Fatality & $\log(\cdot +1)$ & \citet{ucdp} \\
         Months Since last NS Fatality & $\log(\cdot +1)$ & \citet{ucdp} \\
         SB Fatality in $t-s$ &  Identity & \citet{ucdp} \\
         OS Fatality in $t-s$ &  Identity & \citet{ucdp} \\
         NS Fatality in $t-s$ &  Identity & \citet{ucdp} \\
         SB Fatality Count in $t-s$ &  $\log(\cdot +1)$ & \citet{ucdp} \\
         Military Expenditure &  $\log(\cdot)$ & \citet{SIPRI}  \\
         Distance to Capital (CD)  & Identity & \citet{Weidmann}  \\
         Acute Import of MCW (AC) & $\log(\cdot +1)$ & \citet{sipridata2017}\\
         Long-Term Import of MCW (LT) & $\log(\cdot +1)$ & \citet{sipridata2017}\\
         Polity IV Index  &Identity & \citet{marshall2017}  \\
         GDP (Country)  & $\log(\cdot)$ &  \citet{Hegre2019}$^*$  \\
         GDP (Prio-Grid)  & $\log(\cdot)$ & \citet{Hegre2019}$^*$    \\
         Population (Country)  & $\log(\cdot)$ & \citet{Hegre2019}$^*$   \\
         Population (Prio-Grid)  &$\log(\cdot)$ & \citet{Hegre2019}$^*$    \\
         Nightlights &  Identity & \href{https://ngdc.noaa.gov/eog/dmsp/downloadV4composites.html}{NOAA} \\
         Infant Mortality Rate &  Identity  & \href{https://ngdc.noaa.gov/eog/dmsp/downloadV4composites.html}{NASA}  \\
    \end{tabular}
    \caption{Sources of all used variables in our hierarchical hurdle regression with delay structure $s$. * Data was only compiled from different sources by \citet{Hegre2019}, the precise sources are given in the respective appendix.  }
    \label{tab:my_label}
\end{table}
\FloatBarrier

\section{Estimates}
\FloatBarrier

We next present the entire set of estimates from the Hierarchical Hurdle Regression model. For reasons of space, we focused on only a subset of these results in the main paper, here we also show other parametric estimates as well as the non-parametric smooth effects and spatial effects. 

\FloatBarrier
\subsection{Linear Effects}
\FloatBarrier
\begin{table}[t!]
\centering \small
\begin{tabular}{l|l|rrrr} 
 &Covariate & Estimate & Std. Error & t-value & p-value \\    \hline
  \multirow{11}{*}{\begin{sideways}Stage 1\end{sideways}}  & Intercept &  -3.358 &   0.77 &  -4.361 &    $<$ 0.0001 \\
  & NS Fatality in last month  & 0.0951 & 0.1052 & 0.9035 & 0.3663 \\ 
  & OS Fatality in last month & 0.6241 & 0.0804 & 7.7655 & $<$ 0.0001 \\ 
 & SB Fatality in last month & 0.9503 & 0.1000 & 9.5045 & $<$ 0.0001 \\ 
 & SB Fatality Count in last month  & 0.1653 & 0.0288 & 5.7317 & $<$ 0.0001 \\ 
 & Population & 0.3675 & 0.1193 & 3.0820 & 0.0021 \\ 
& GDP & -0.2456 & 0.0875 & -2.8057 & 0.005 \\ 
  &Polity IV Index & -0.0105 & 0.0103 & -1.0129 & 0.3111 \\ 
 & Military Expenditure & 0.0624 & 0.0511 & 1.2227 & 0.2215 \\ 
  &Long-Term Import of MCW & -0.0898 & 0.0452 & -1.9851 & 0.0471\\ 
  &Acute Import of MCW & 0.1104 & 0.0285 & 3.8673 & 0.0001 \\ \hline
 \multirow{16}{*}{\begin{sideways}Stage 2\end{sideways}}  & Intercept &  -7.194 & 0.595 & -12.084 &  $<$ 0.0001 \\ 
 & NS Fatality in last month & -0.0256 & 0.1032 & -0.2481 & 0.8041 \\ 
 & OS Fatality in last month & 0.5672 & 0.0461 & 12.3147 & $<$ 0.0001 \\ 
  & SB Fatality in last month & 0.8287 & 0.0603 & 13.7342 & $<$ 0.0001 \\ 
 & Nightlights  & 2.4335 & 0.2054 & 11.8460 & $<$ 0.0001 \\ 
 & Infant Mortality Rate & 0.0005 & 0.0001 & 7.2693 & $<$ 0.0001 \\ 
  &SB Fatality Count in last month  & 0.1725 & 0.0236 & 7.3044 & $<$ 0.0001 \\ 
 & Population  & -0.1016 & 0.0339 & -3.0006 & 0.0027 \\ 
  & GDP  & -0.0096 & 0.0251 & -0.3829 & 0.7018\\ 
 & Polity IV Index & -0.0434 & 0.0043 & -10.0533 &  $<$ 0.0001 \\ 
  & Military Expenditure & 0.0923 & 0.0173 & 5.3437 & $<$ 0.0001 \\ 
 & Distance to Capital (CD)  & 0.0001 & 0.0001 & 1.2348 & 0.2169 \\ 
 & Long-Term Import of MCW (LT)  & -0.0291 & 0.0229 & -1.2678 & 0.2049\\ 
 & Acute Import of MCW (AC)  & -0.0643 & 0.0179 & -3.5986 & 0.0003 \\ 
 & LT×CD & -0.0002 & 0.0000 & -9.5322 & $<$ 0.0001 \\ 
 & AC×CD & 0.0000 & 0.0000 & 1.4582 & 0.1448 \\ 
   \hline
  \multirow{13}{*}{\begin{sideways}Stage 3\end{sideways}}   & Intercept & 
  7.221 &  1.416 &   5.1 &    $<$ 0.0001 \\
  & Infant Mortality Rate & -0.0013 & 0.0002 & -7.3465 & $<$ 0.0001 \\ 
   & OS Fatality in last month & 0.0354 & 0.0383 & 0.9249 & 0.3551 \\ 
   & NS Fatality in last month & -0.0097 & 0.0928 & -0.1050 & 0.9164 \\ 
   & Population & 0.0782 & 0.0692 & 1.1302 & 0.2584\\ 
   & GDP & -0.0315 & 0.0567 & -0.5554 & 0.5786\\ 
   & Polity IV Index & -0.0357 & 0.0118 & -3.0304 & 0.0024 \\ 
  &  Military Expenditure  & 0.1870 & 0.0438 & 4.2712 &   $<$ 0.0001\\ 
  & Distance to Capital (CD) & 0.0020 & 0.0002 & 11.2872 & $<$ 0.0001 \\ 
  & Long-Term Import of MCW (LT) & 0.6844 & 0.0519 & 13.1837 & $<$ 0.0001 \\ 
 &  Acute Import of MCW (AC)  & -0.0140 & 0.0418 & -0.3349 & 0.7377\\ 
  & LT×CD & -0.0009 & 0.0001 & -12.7715 &  $<$ 0.0001 \\ 
 & AC×CD & -0.0002 & 0.0001 & -3.5346 & 0.0004 \\ 
   \hline
\end{tabular}
\caption{Parametric estimates with $s = 2$ and complete training data used for the real forecasts (January 1990 to August 2020).} 
\label{tab.gam}
\end{table}

\begin{figure}[t!]
		\centering
		\includegraphics[width=0.45\linewidth]{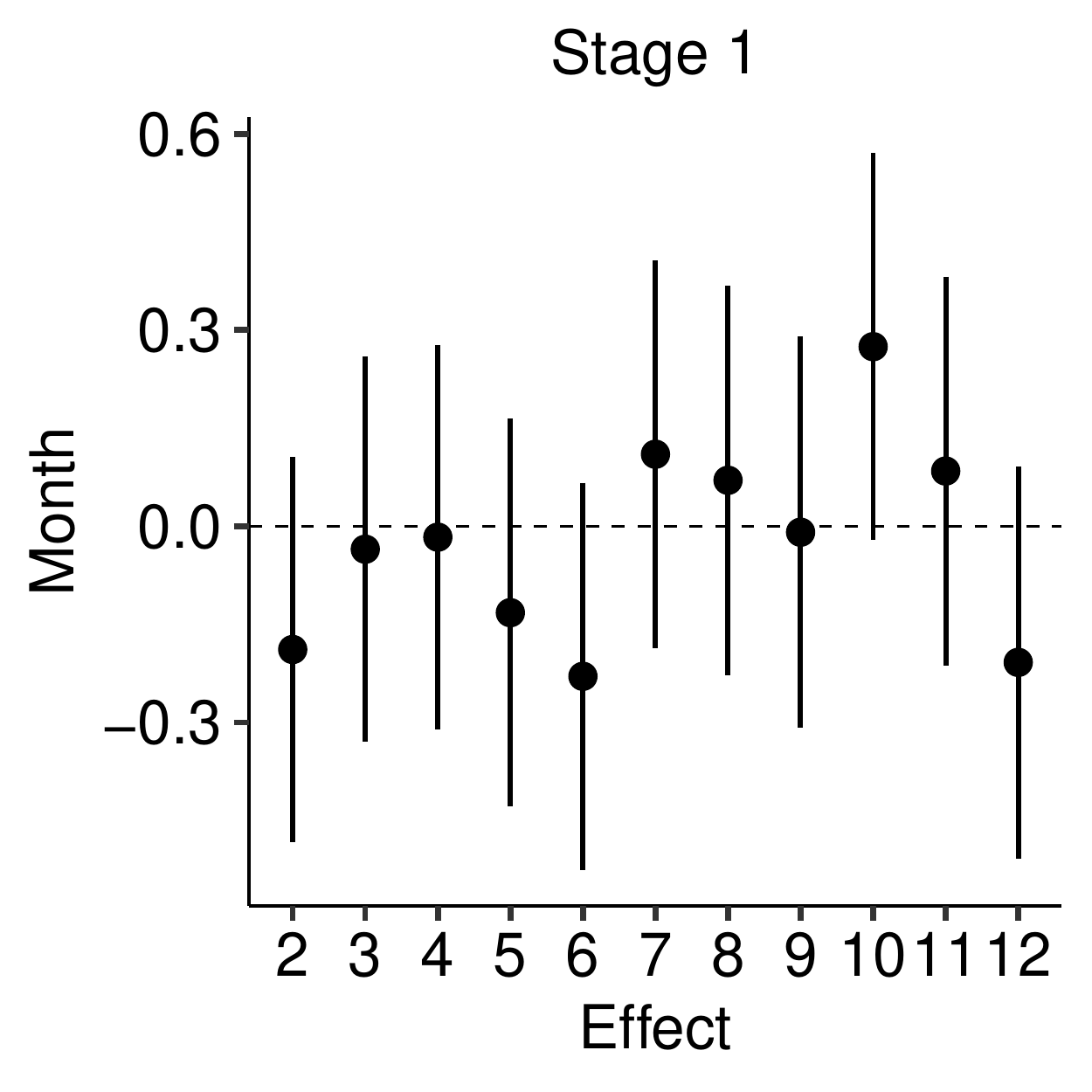}
		\includegraphics[width=0.45\linewidth]{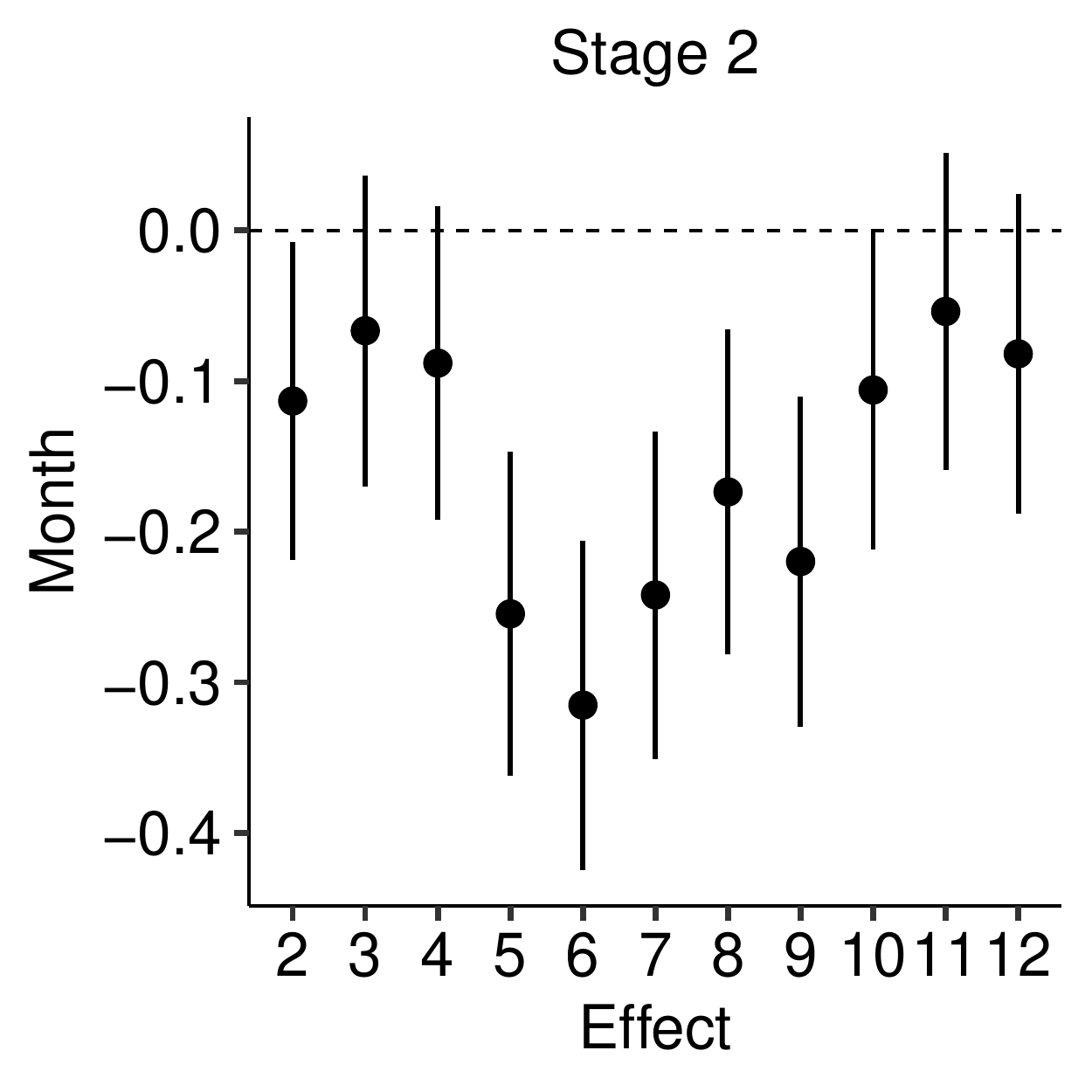}
		\includegraphics[width=0.45\linewidth]{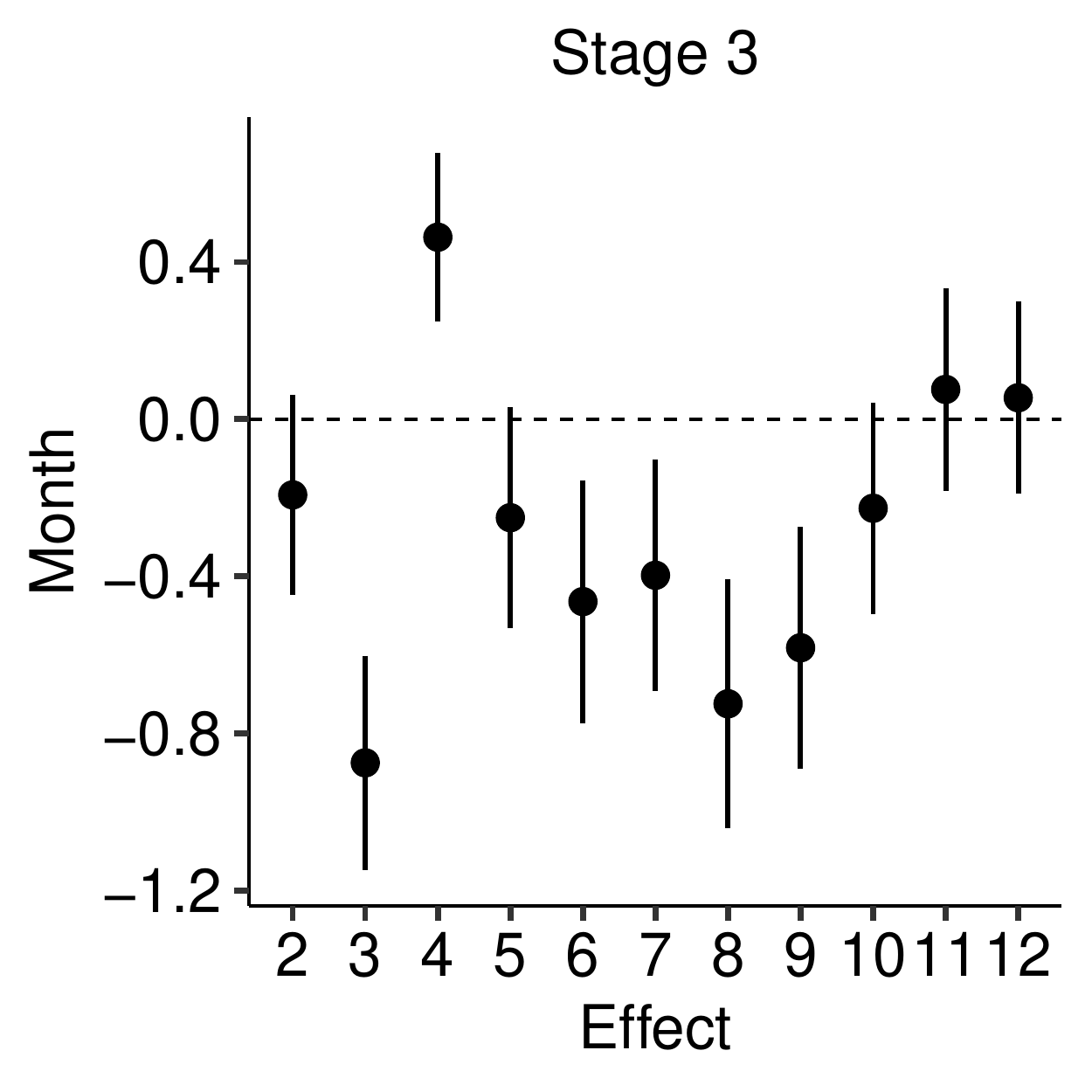}
		\caption{Estimates regarding the monthly dummy variables. In each case the reference class is January.}
				\label{fig:monthly}
\end{figure}

Figure \ref{fig:monthly} presents the estimates for the eleven month-dummies we include in all three stages of the model to capture seasonal effects. Here, it is noteworthy that while seasonality seems to play no decisive role regarding the country-wide incidence of violence, this is different when we examine the grid-level results in stages two and three. The results in Figure \ref{fig:monthly} suggest that both the probability and intensity of fighting is lowest between May and September. Fighting is further expected to de-escalate in March but then escalate in April. These estimates may hint at factors such as seasonal differences in agricultural workload playing a role in the occurrence and intensity of fighting.   

\FloatBarrier
\subsection{Smooth Effects}
\FloatBarrier

\begin{figure}[t!]
		\centering
		\includegraphics[width=0.45\linewidth]{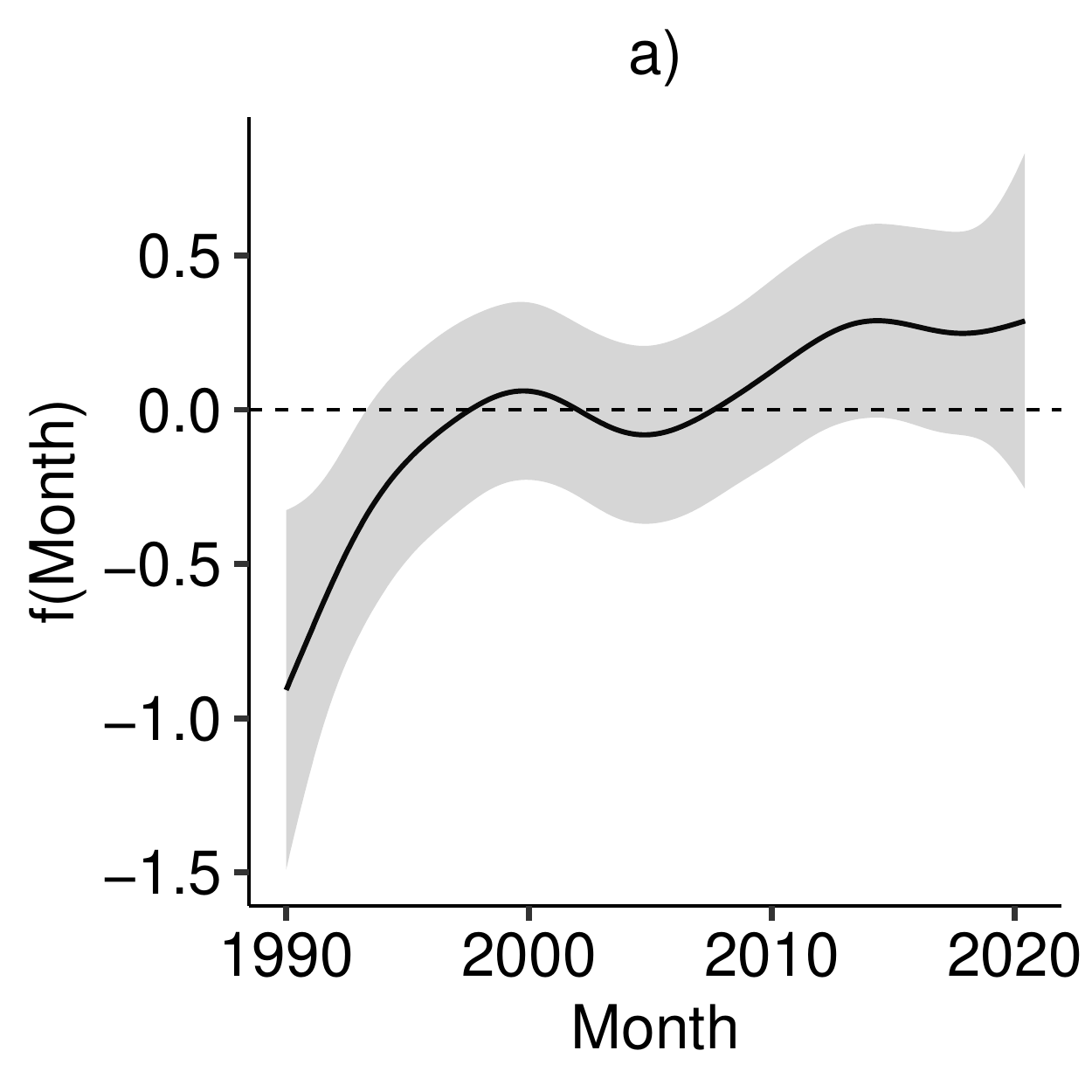}
		\includegraphics[width=0.45\linewidth]{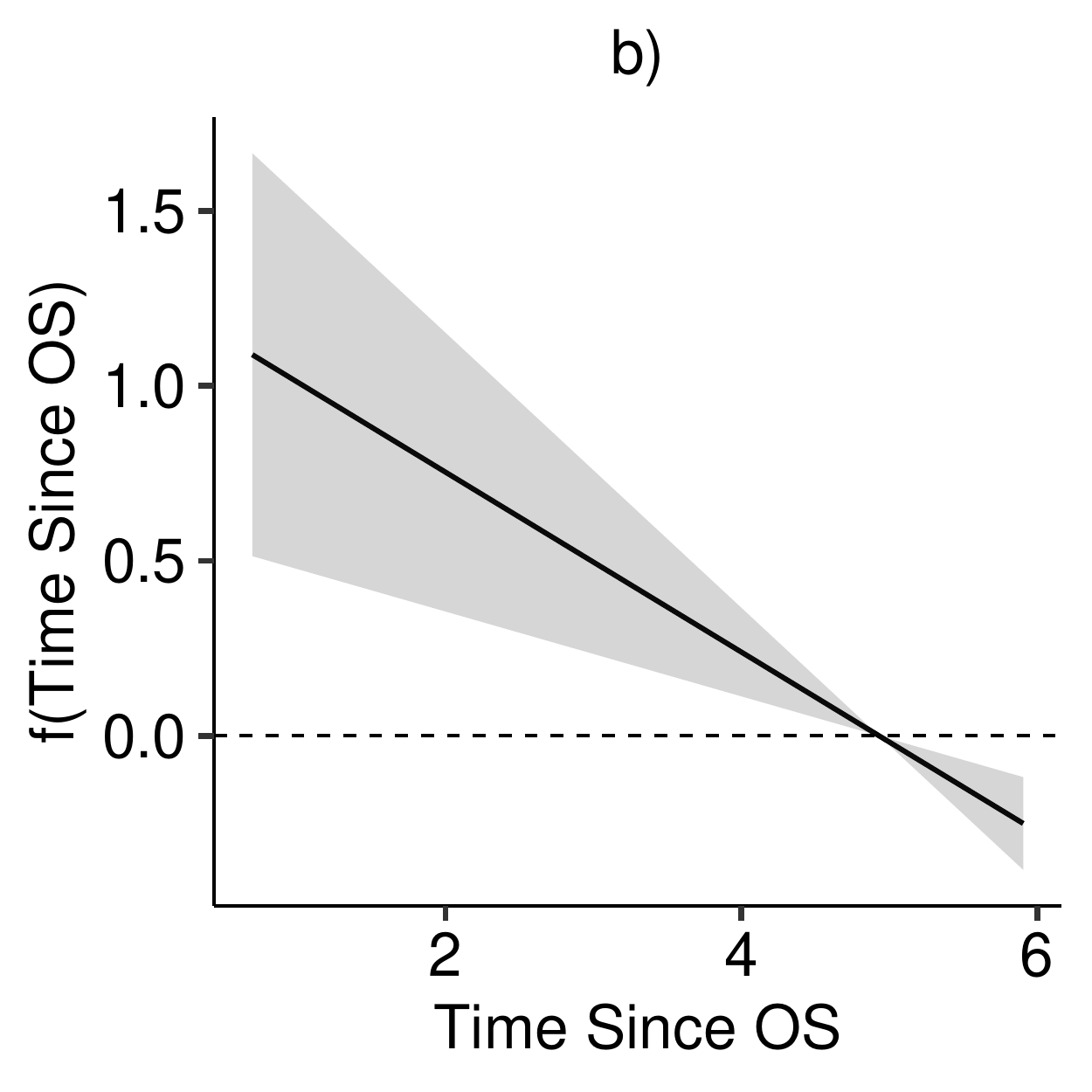}
		\includegraphics[width=0.45\linewidth]{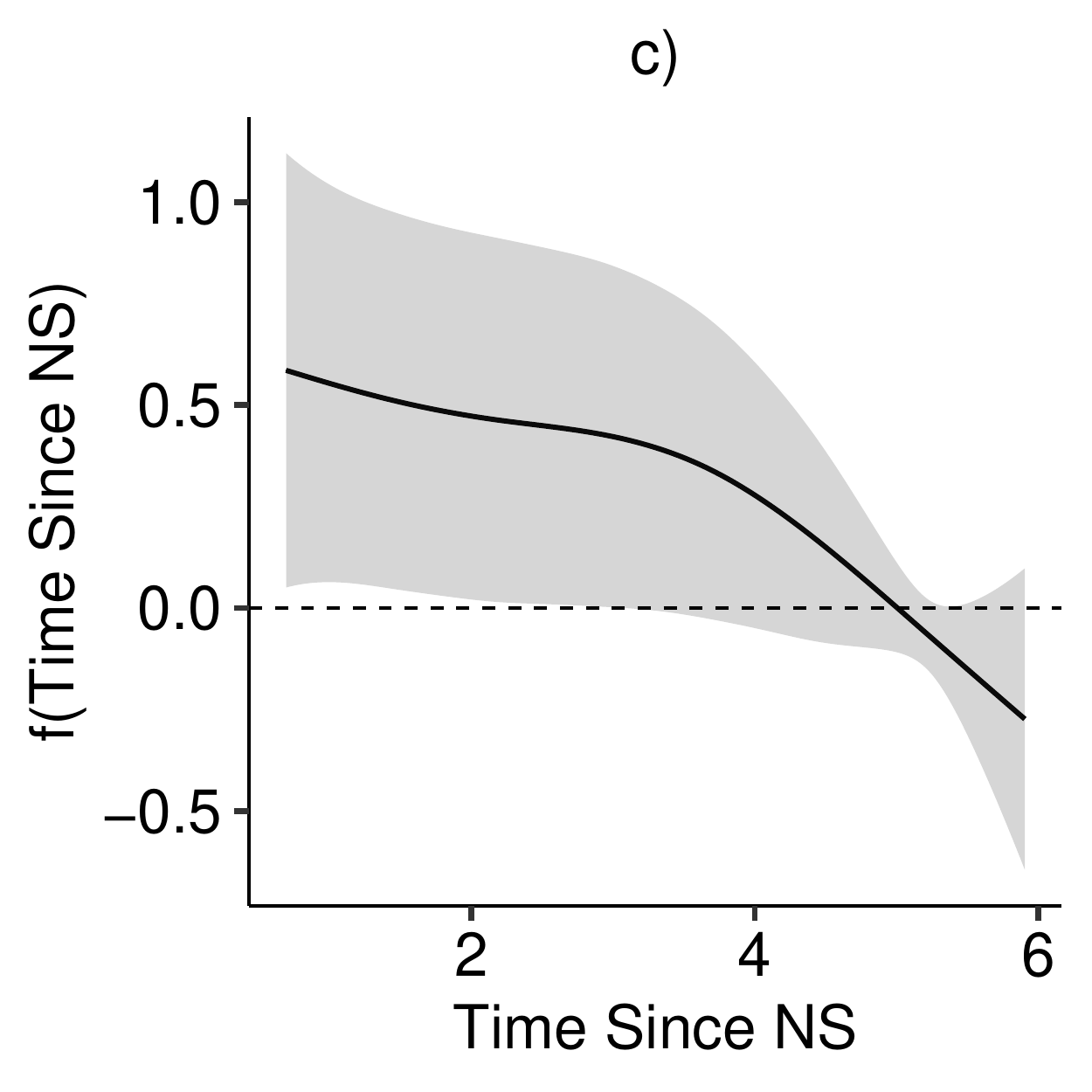}
		\includegraphics[width=0.45\linewidth]{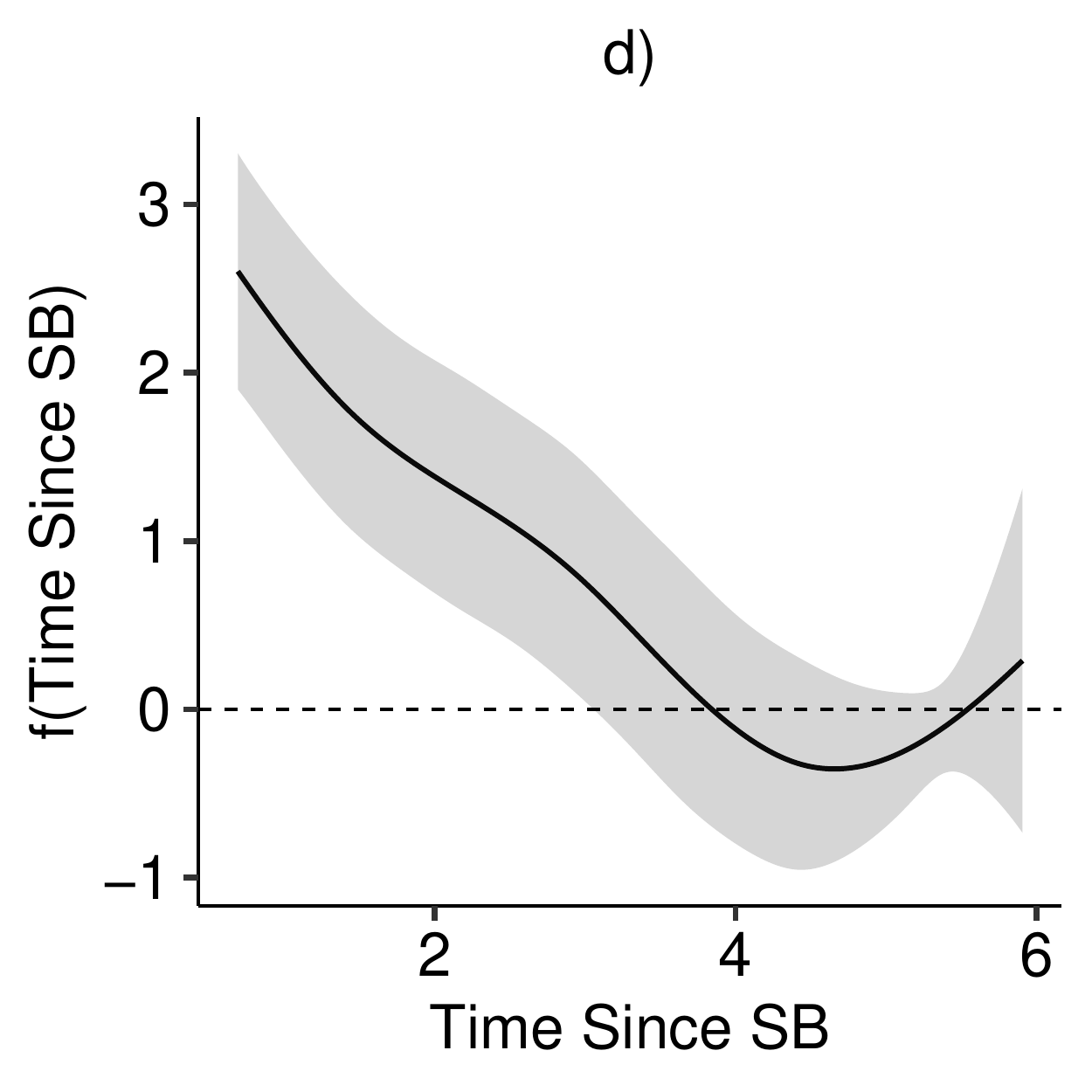}
		\caption{Estimates of smooth effects from the Stage 1 Model.}
				\label{fig:mod_1}
	\end{figure}
\begin{figure}[ht!]
		\centering
		\includegraphics[width=0.45\linewidth]{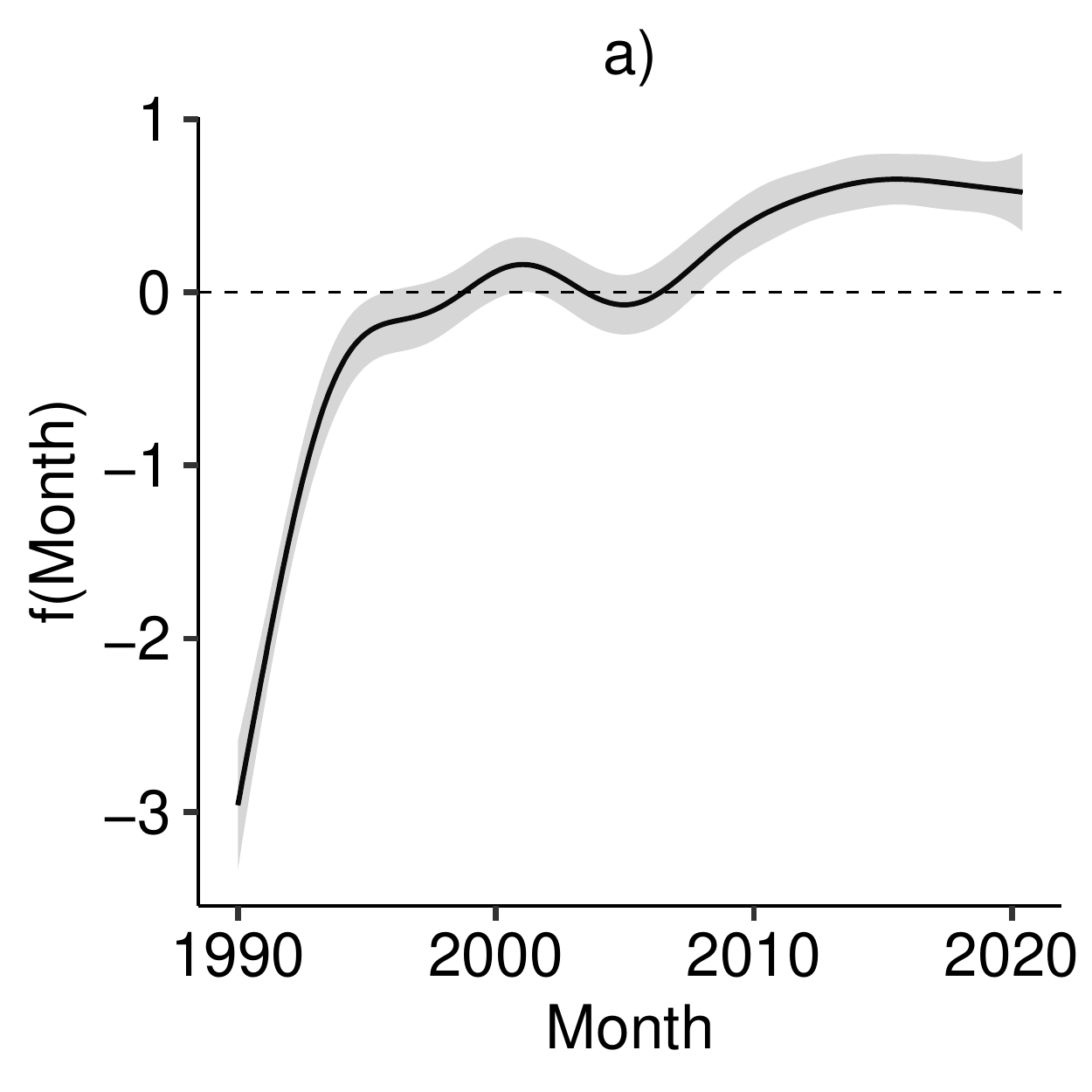}
		\includegraphics[width=0.45\linewidth]{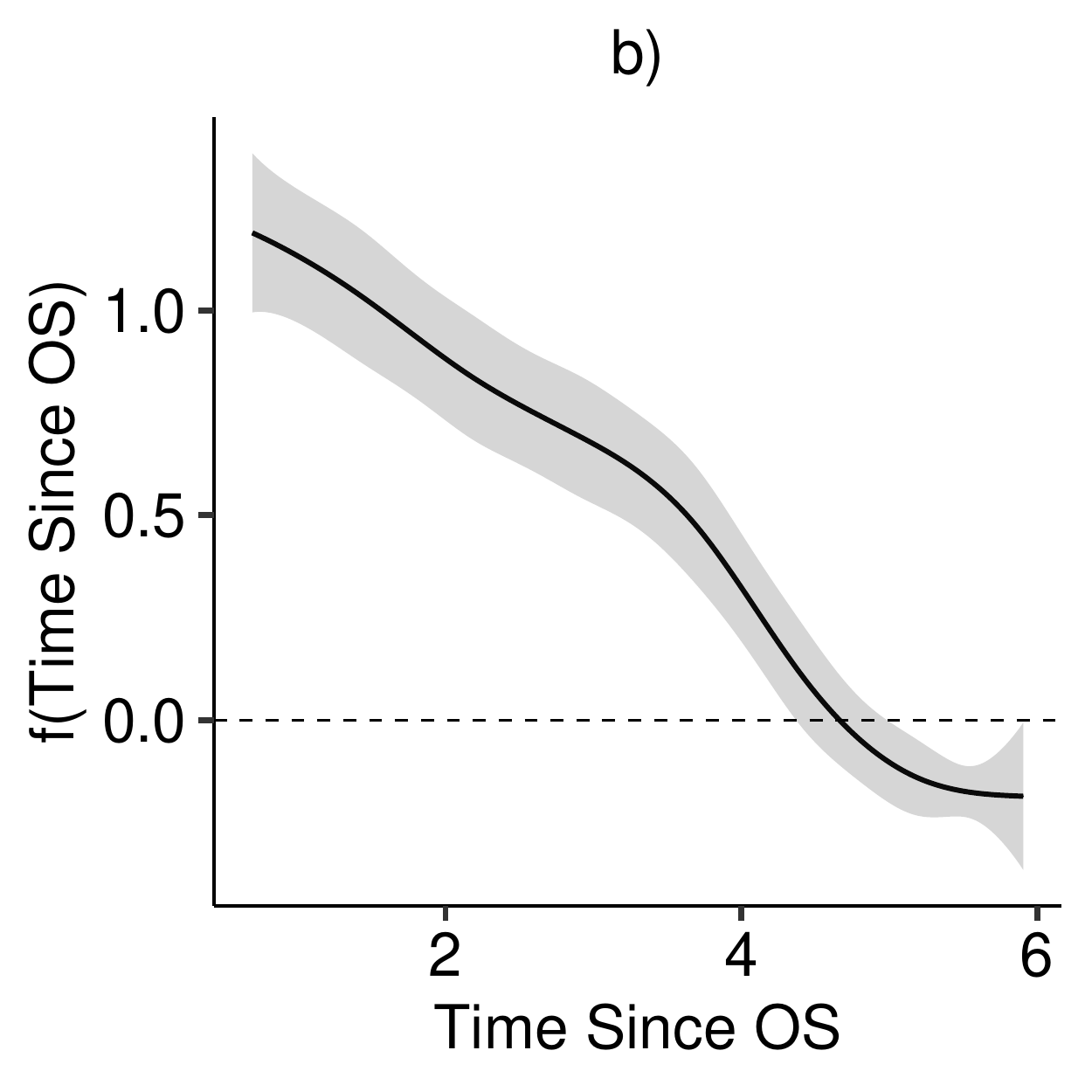}
		\includegraphics[width=0.45\linewidth]{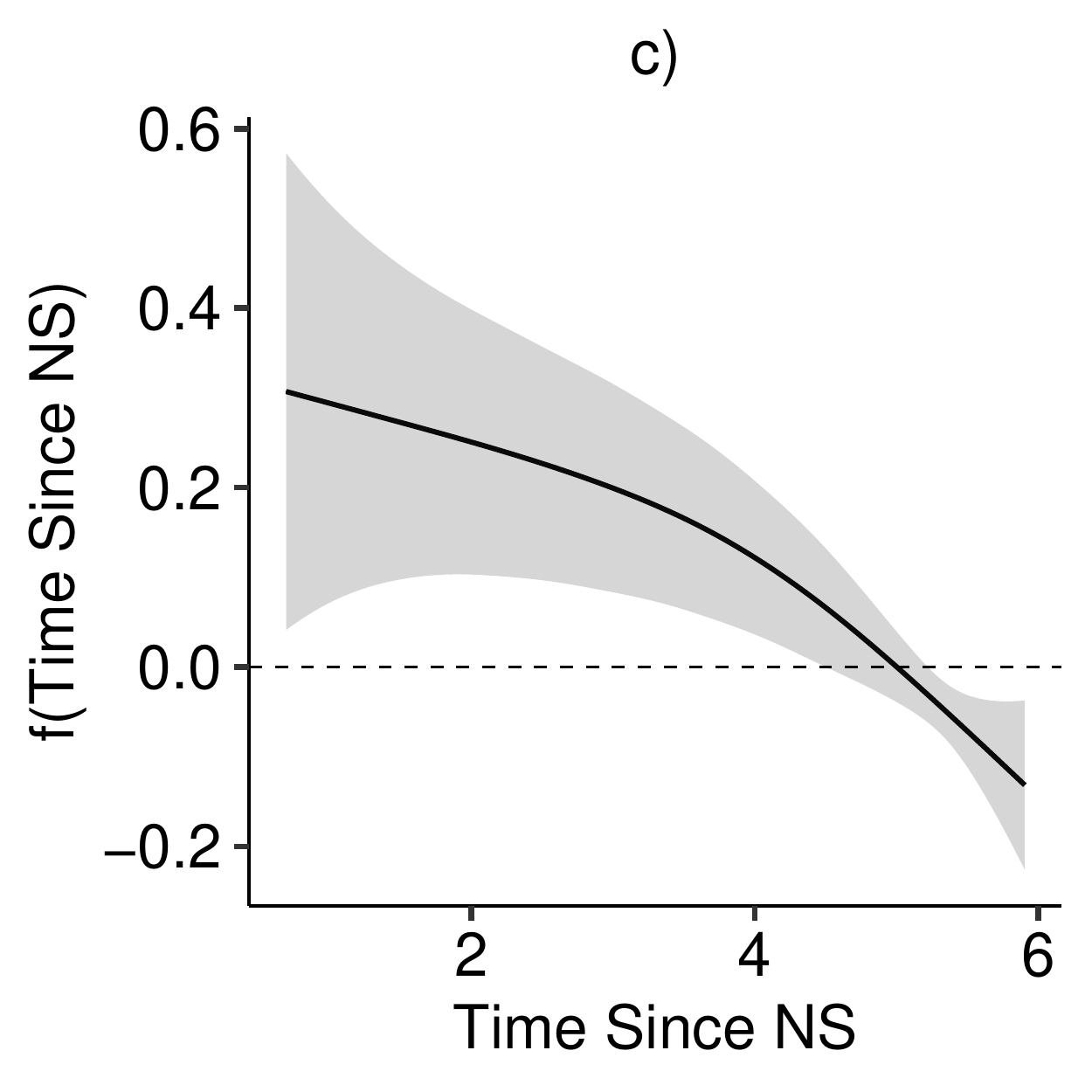}
		\includegraphics[width=0.45\linewidth]{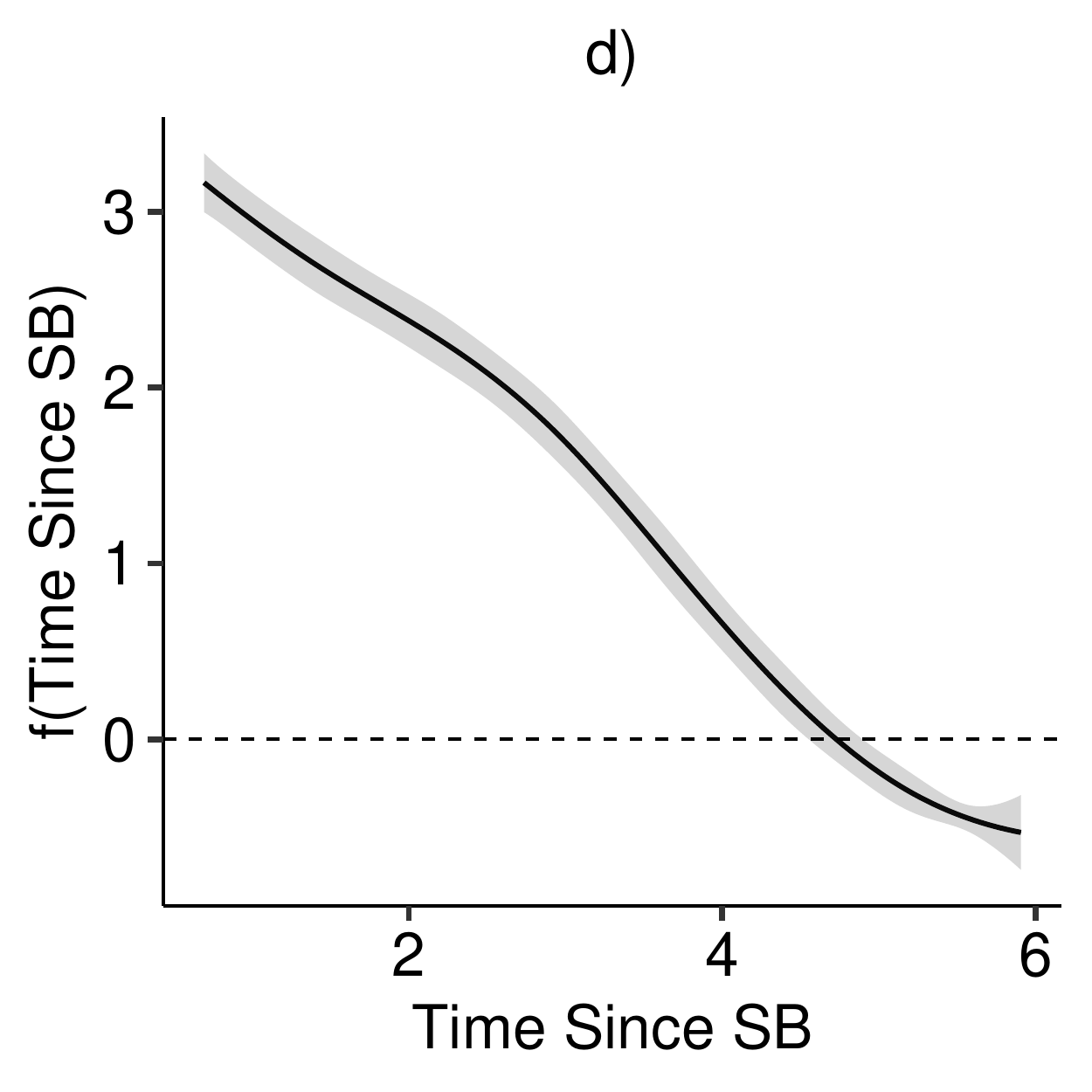}
		\caption{Estimates of smooth effects from the Stage 2 Model.}
				\label{fig:mod_2}
\end{figure}

\begin{figure}[ht!]
		\centering
		\includegraphics[width=0.45\linewidth]{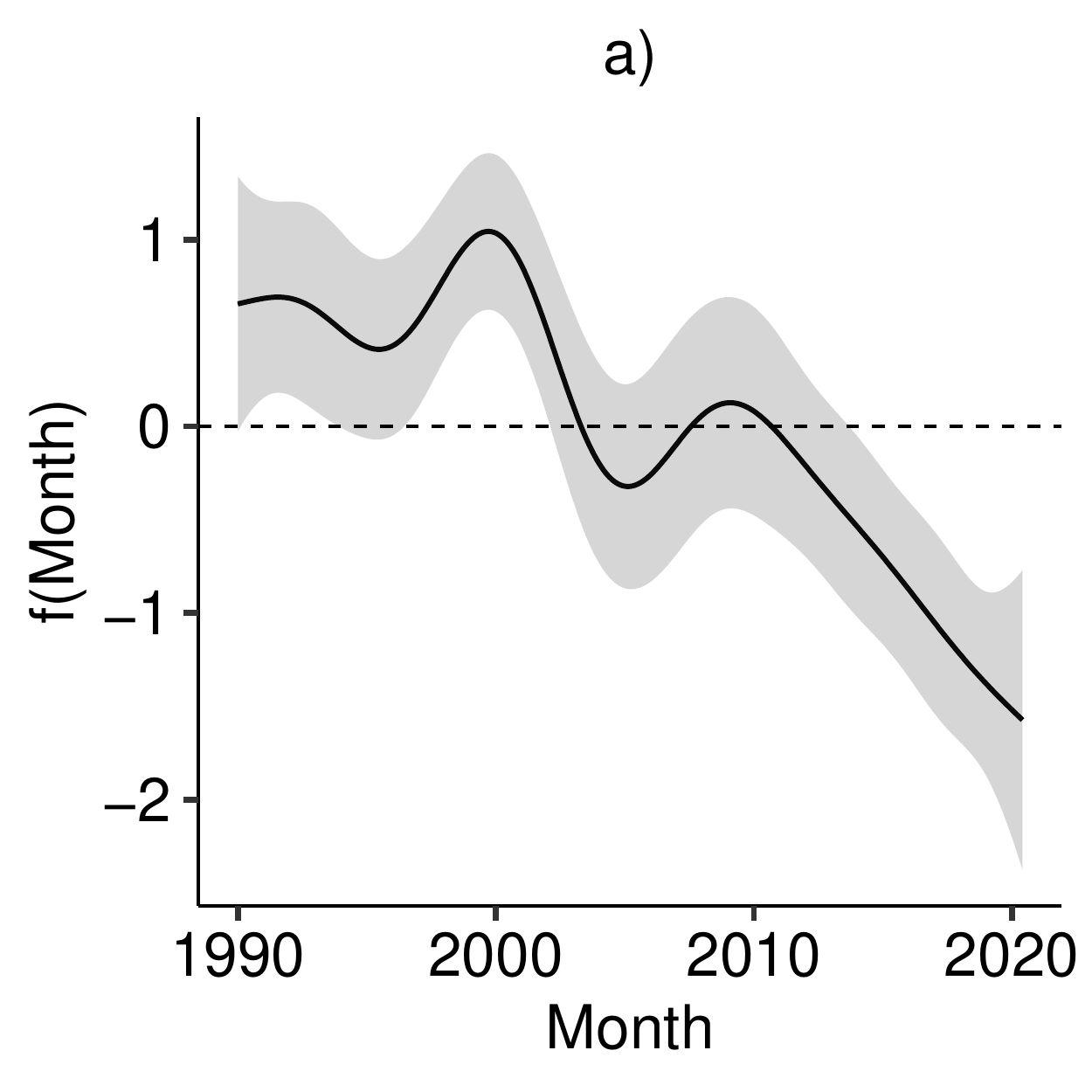}
		\includegraphics[width=0.45\linewidth]{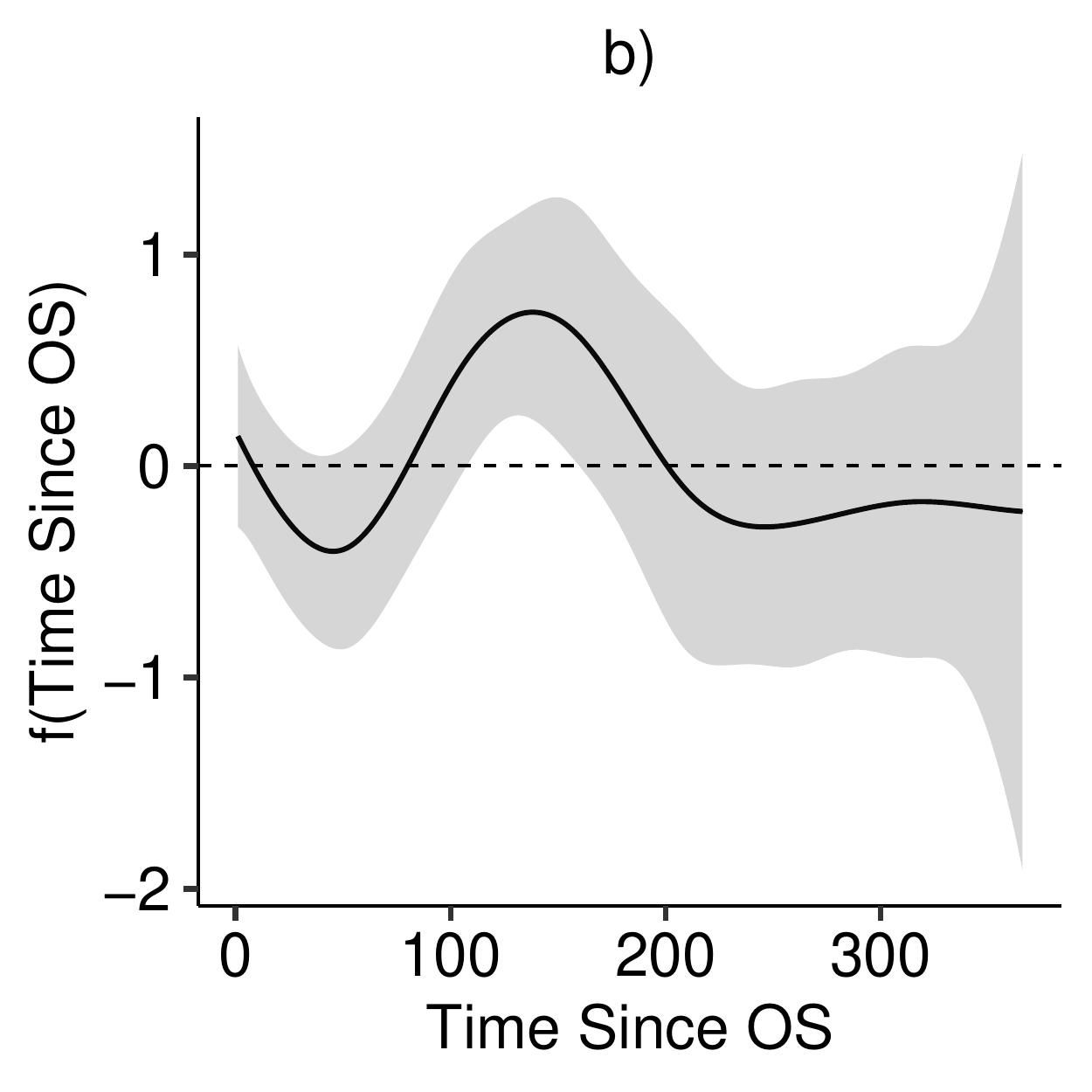}
		\includegraphics[width=0.45\linewidth]{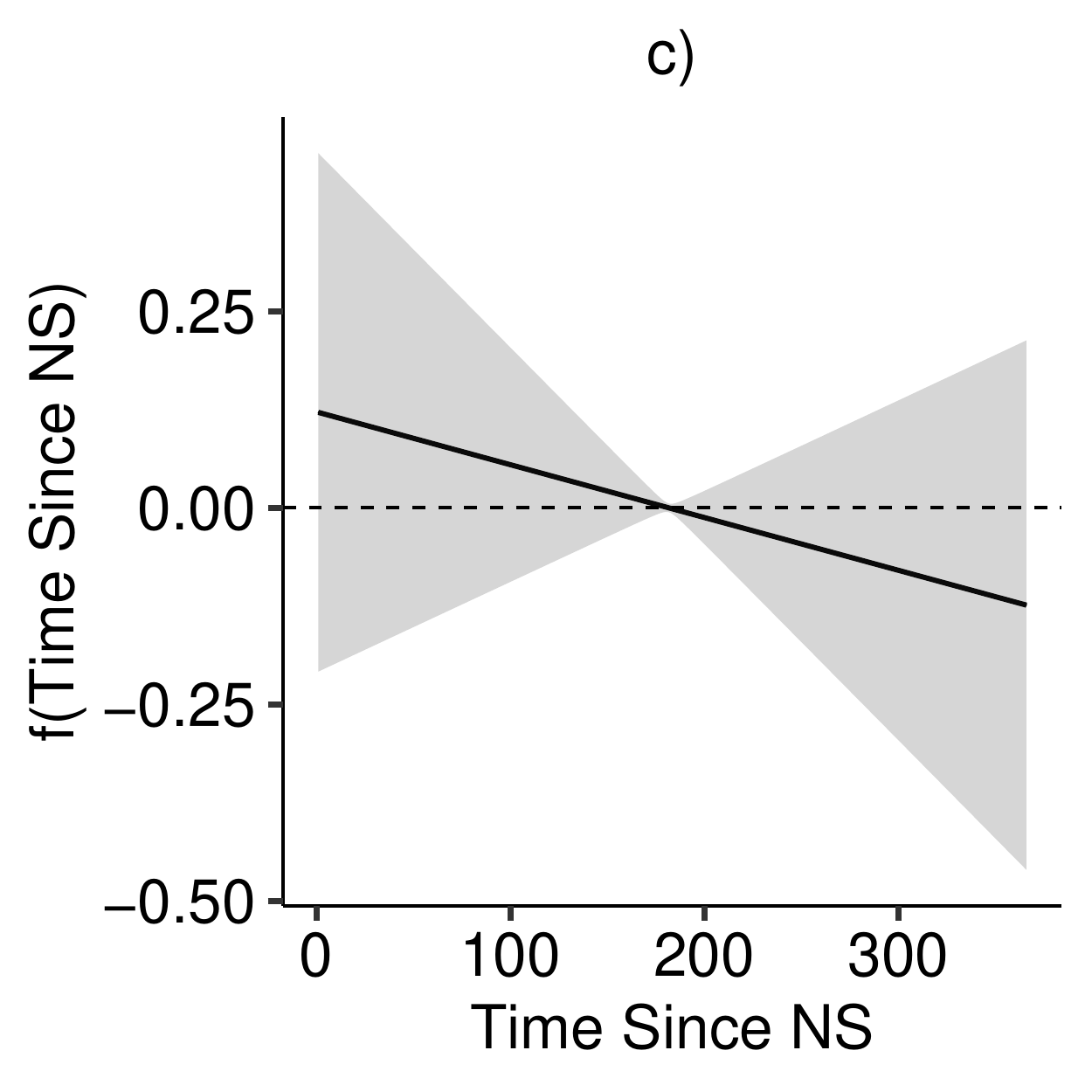}
		\includegraphics[width=0.45\linewidth]{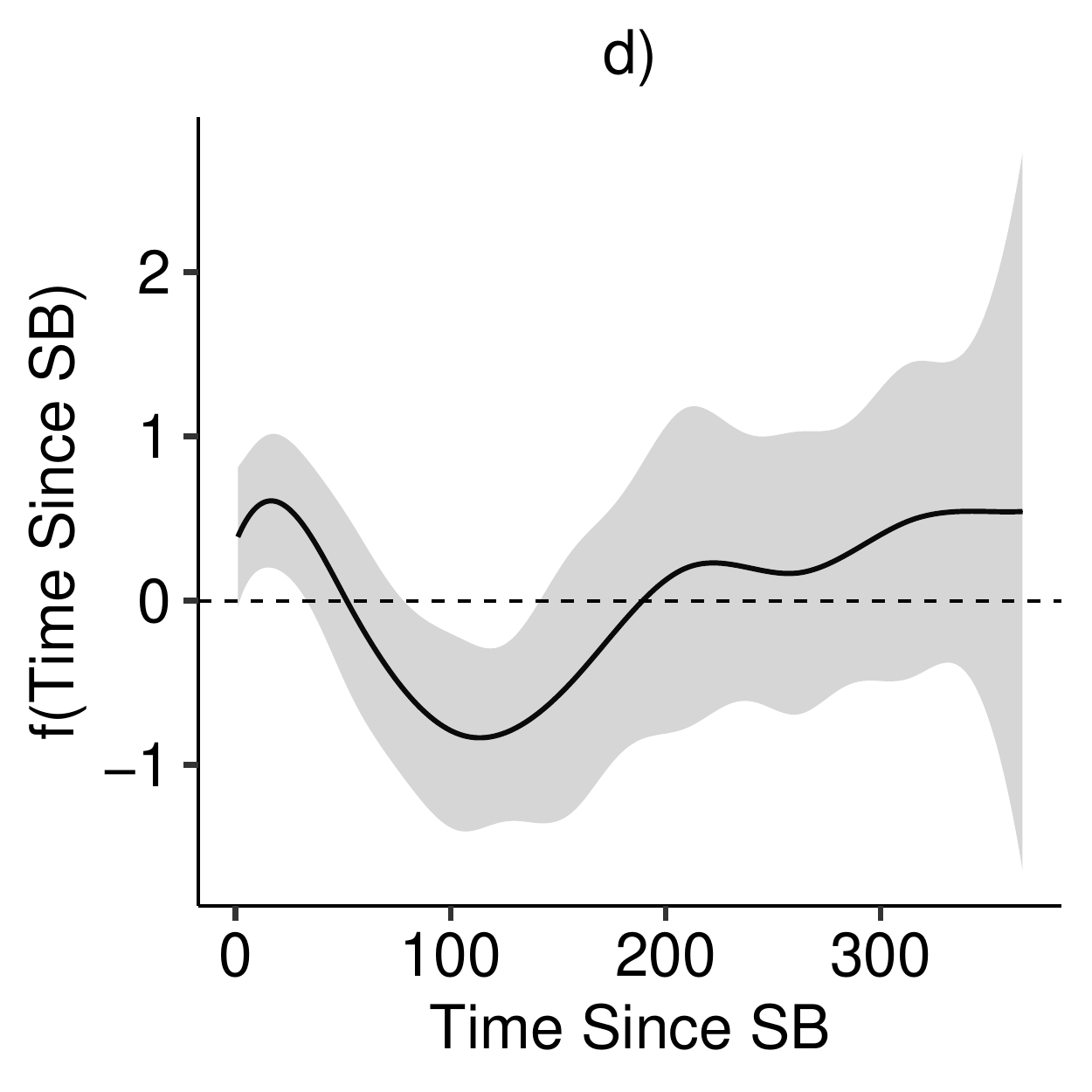}
		\includegraphics[width=0.45\linewidth]{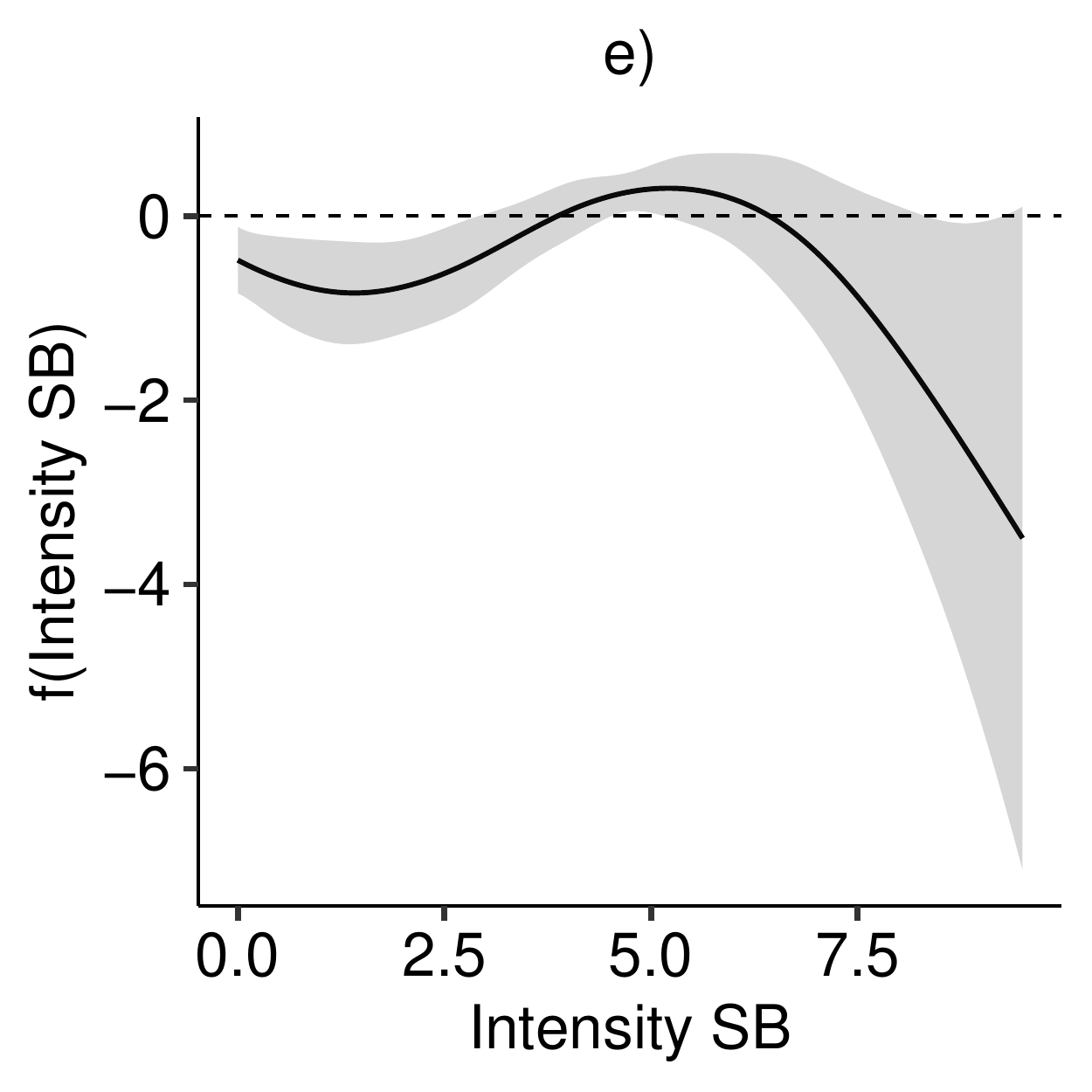}
		\caption{Estimates of smooth effects from the Stage 3 Model.}
				\label{fig:mod_3}
\end{figure}

The estimated smooth effects are depicted in Figures \ref{fig:mod_1} to \ref{fig:mod_3}. Coincidentally, the effects of the time since any conflict (OS, NS, SB) all share a similar monotonically decreasing form. As it turns out, this shape is reminiscent of the exponential decay function, which is applied to transform those covariates in \citet{Hegre2019}. Thus our results confirm that this type of effect is suitable and recovered in our flexible approach.   
\newpage
~ 
\newpage
\FloatBarrier

\subsection{Random and Spatial Effects}
\FloatBarrier

\begin{figure}[t!]
		\centering
		\includegraphics[width=0.45\linewidth, page = 1,trim={2.5cm 0cm 2.5cm 0cm},clip]{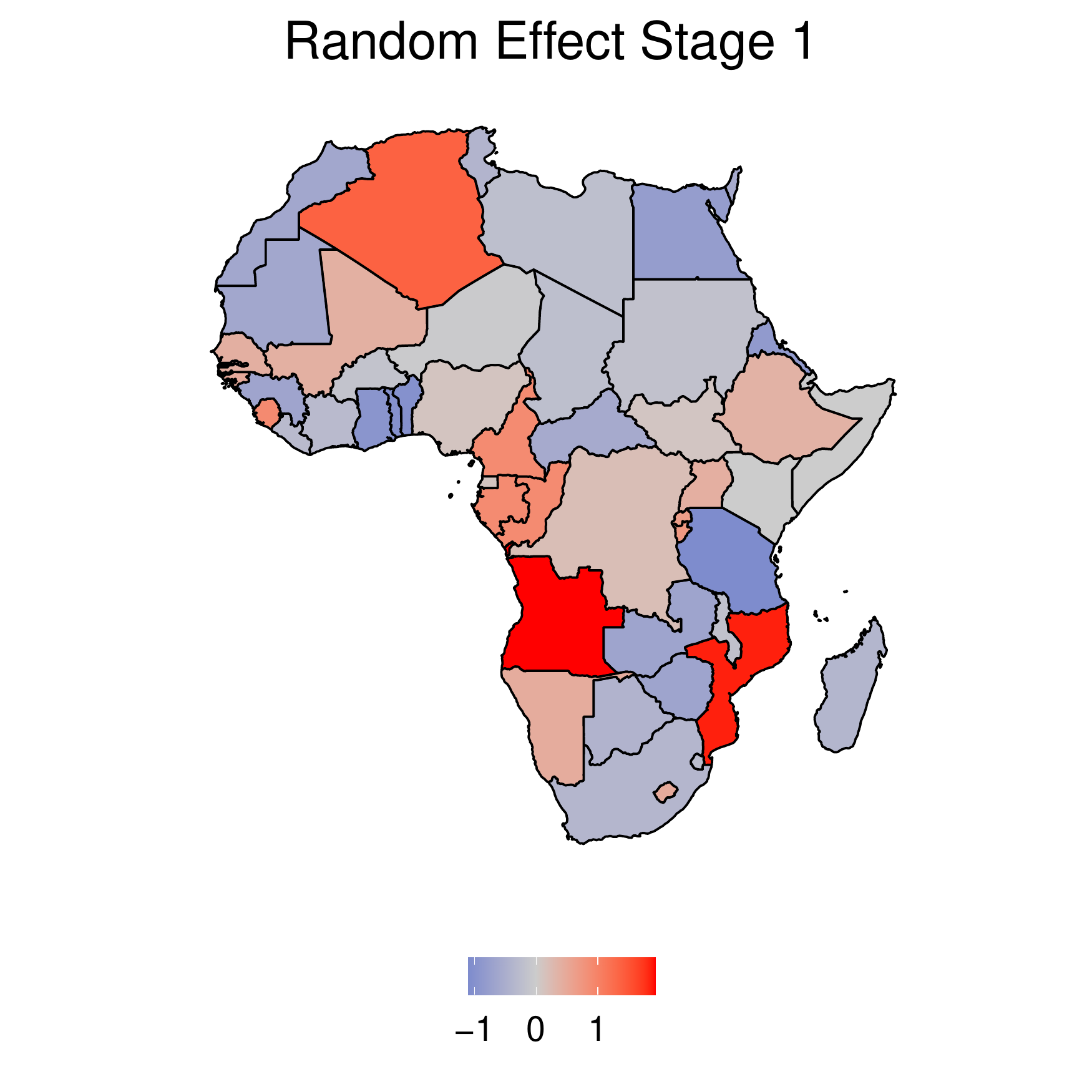}
		\includegraphics[width=0.45\linewidth, page = 2,trim={2.5cm 0cm 2.5cm 0cm},clip]{random_spatial_effects.pdf}
		\caption{Random (left) and spatial (right) effect of the model at stage 1.}
		\label{fig:randomSpat}
\end{figure}

\begin{figure}[t!]
		\centering
		\includegraphics[width=0.45\linewidth, page = 3,trim={2.5cm 0cm 2.5cm 0cm},clip]{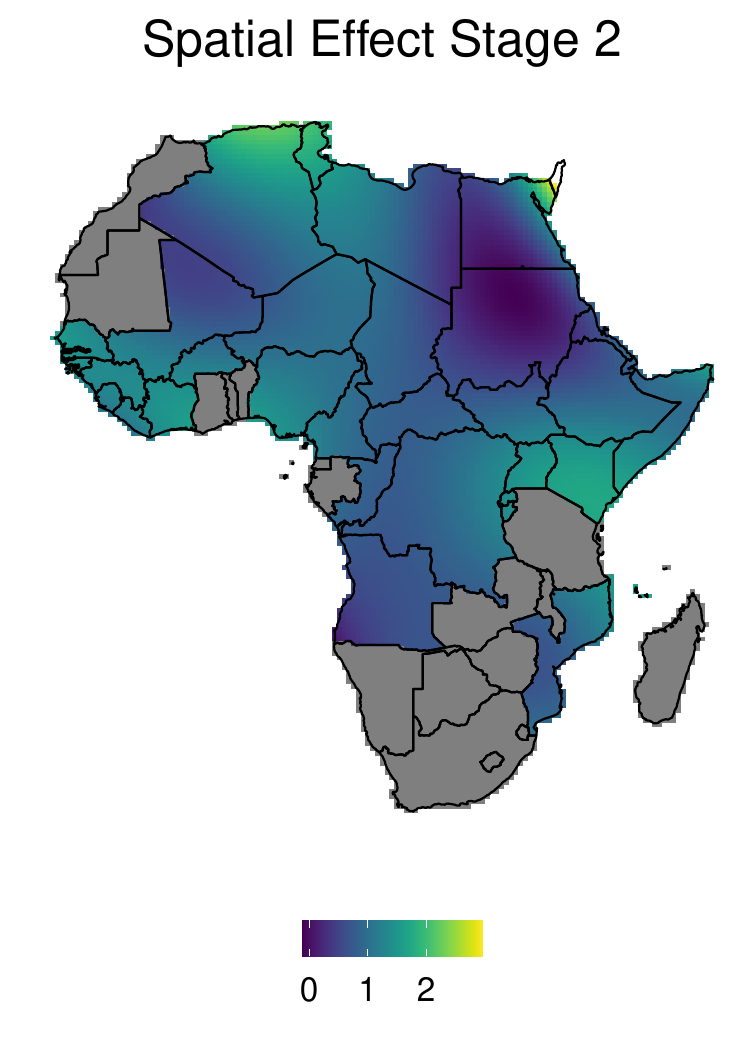}
		\includegraphics[width=0.45\linewidth, page = 4,trim={2.5cm 0cm 2.5cm 0cm},clip]{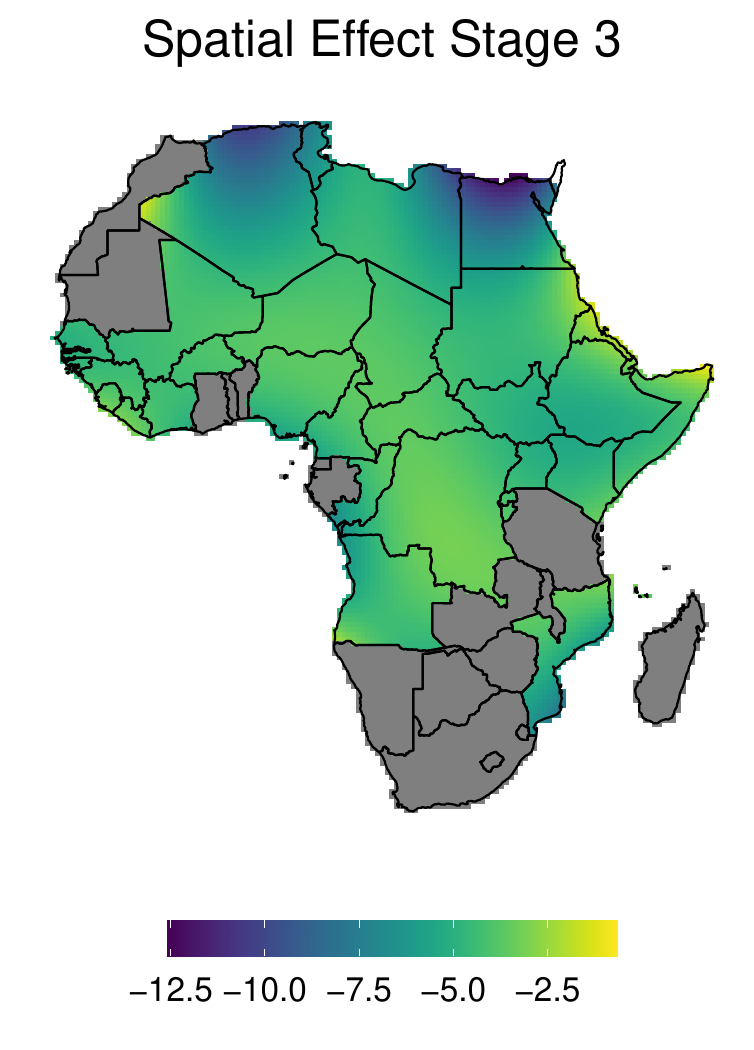}
		\caption{Fixed spatial effects of the model at stage 2 (left) and 3 (right). Data is only shown for countries that had least once more than 25 fatalities within a year.}
\end{figure}

\begin{figure}[ht!]
		\centering
		\includegraphics[width=0.8\linewidth]{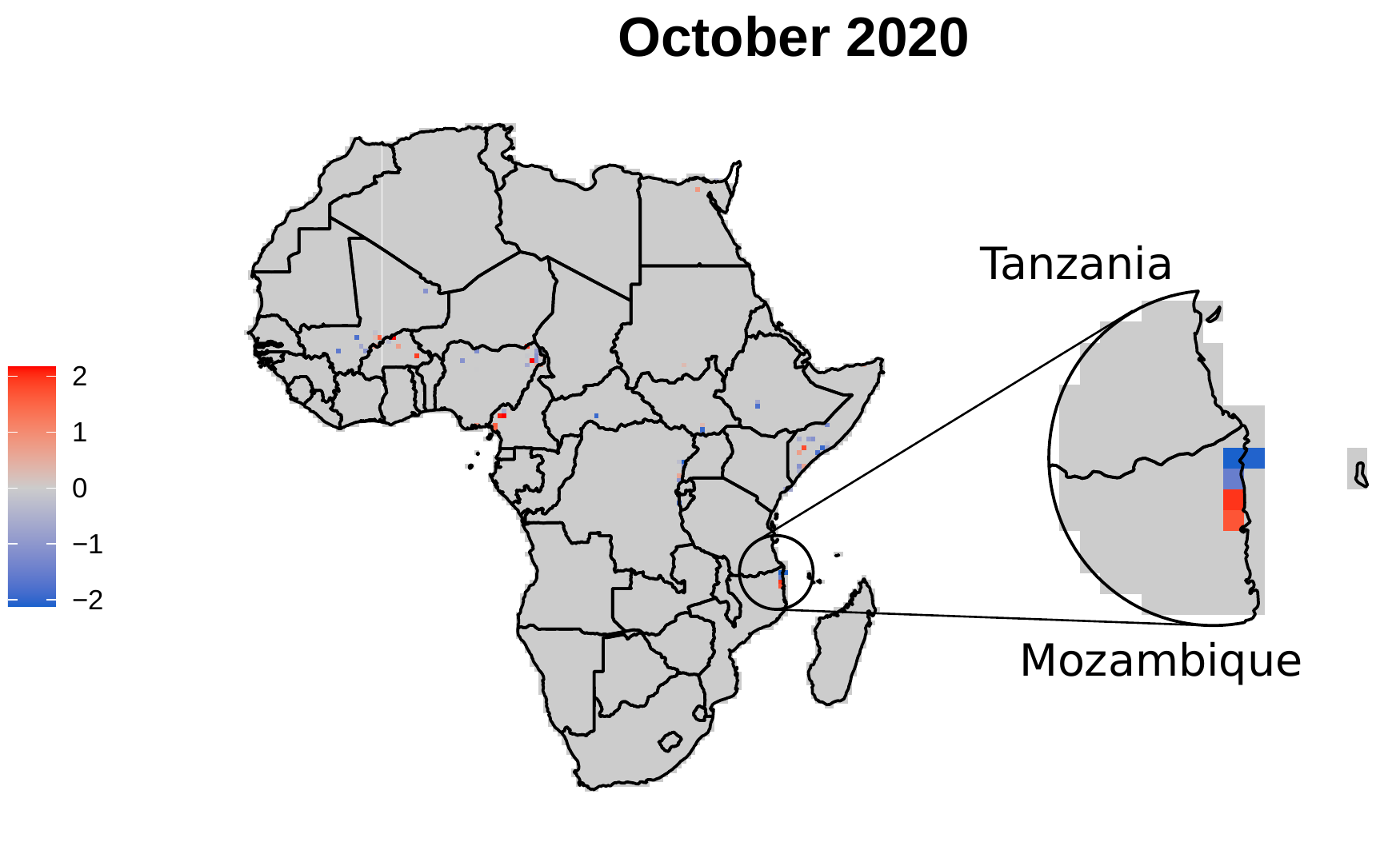}
		\includegraphics[width=0.8\linewidth]{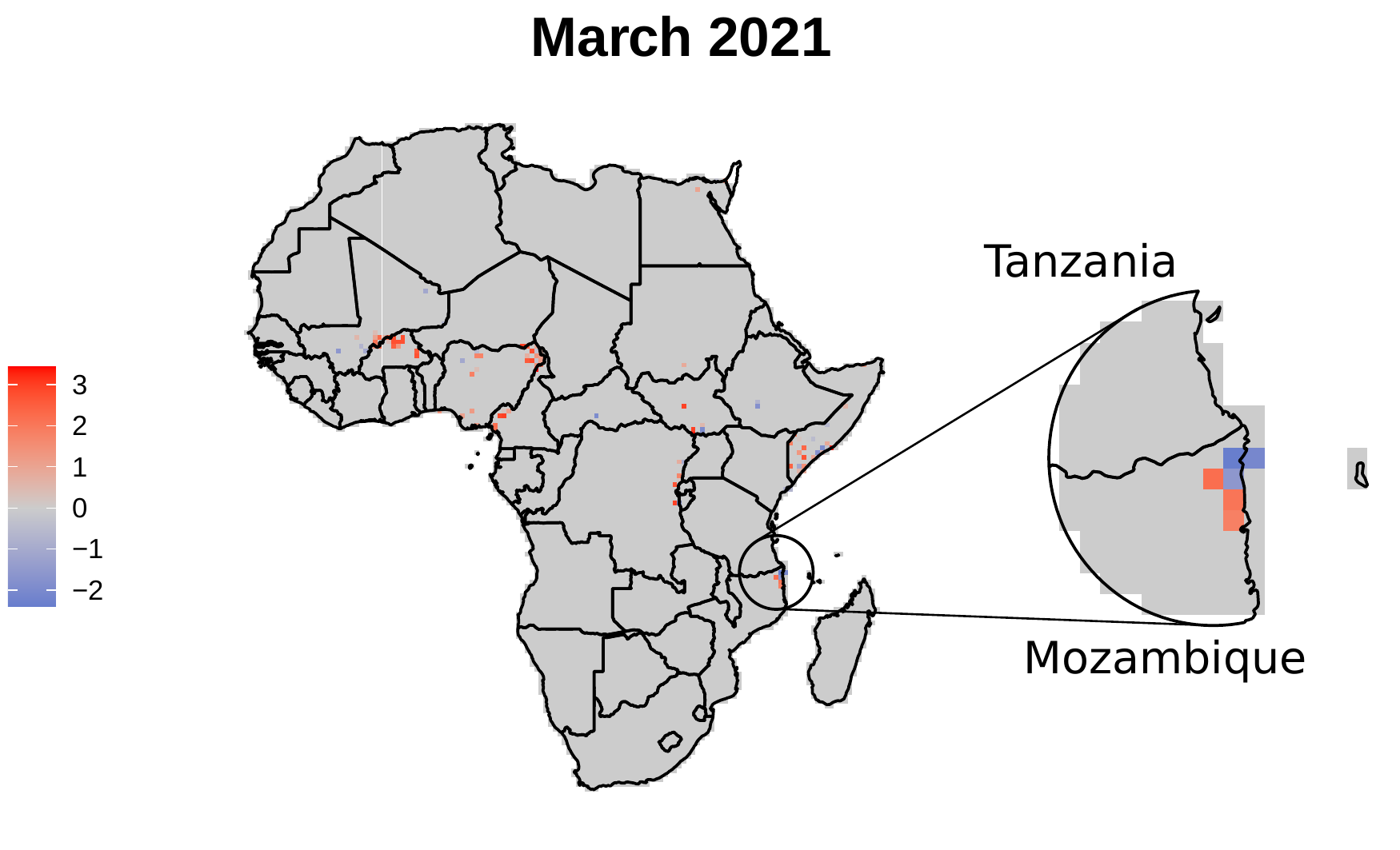}
		\caption{Forecasted changes in fatalities to October 2020 and March 2021 focused on the boarder region between Mozambique and Tanzania.}
		\label{fig:forecast}
	\end{figure}

As the model's final component, we present the random and fixed spatial effects included in the hierarchical hurdle regression model. Most importantly, the random effect in stage one allows us to capture latent differences between countries, which are not accounted for by our covariates but make some countries more and others less violent. As such, the left panel in figure \ref{fig:randomSpat} identifies Angola and Mozambique as particularly prone to experience state-based conflict, whereas violence is found to be unlikely in countries such as Tanzania, Ghana, Togo, and Benin. This result matches our forecast for the potential escalation of violence emanating from the Islamist insurgency in Mozambique's Cabo Delgado province (see figure \ref{fig:forecast}) as violence is predicted to move southwards, approaching the province's capital, but not spill-over into Tanzania.

\FloatBarrier

\section{Forecasts in Mozambique and Tanzania}
\label{sec:real}
\FloatBarrier

\clearpage

\clearpage
\section{Implementation}

All scripts used for obtaining the forecasts, estimating the models, and producing the visualisations are written in $\mathtt{R} $ \citep{R} and available from the journal's replication data page. 

Since some of the covariates are not available throughout the entire time of the training data, we use the $\mathtt{R}$-package $\mathtt{Amelia}$ for one imputation of the complete dataset \citep{amelia}.  
Conditional on having imputed all missing observations,  estimation of the parameters $\theta^{(1)}, \theta^{(2)},$ and $\theta^{(3)}$ we to carry out three separate routines. For optimising each stage-wise penalised likelihood, we leverage the fast computational routines implemented in the $\mathtt{R}$ packages $\mathtt{mgcv}$ and $\mathtt{countreg}$ \citep{wood2017,Zeileis2008}. 

In order to tune the threshold parameter for classification tasks, \citet{Sheng2006} use a $K$-fold cross-validation scheme that evaluates the loss function at all unique predicted probabilities to minimise loss defined in (7) for each fold. Because we have to tune more than one threshold, the so-called \textsl{curse of dimensionality} results in exponentially more candidate values at which we have to evaluate the loss function. To minimise the loss defined in (7) of the main article in a time-efficient fashion, we hence utilise the differential evolutionary algorithm for global optimisation implemented in the $\mathtt{R}$-package $\mathtt{DEoptim}$ \citep{Mullen2011}.

The visualisations including maps were obtained with the help of $\mathtt{ggplot2, sf}$ and $\mathtt{cowplot}$ \citep{wickham2016,cowplot,sf}. Due to the large sizes of objects data management was carried out with  $\mathtt{data.table, lubridate, tidyverse}$ and $ \mathtt{pryr}$ \citep{lubridate,datatable,pryrm,tidyverse}. Lastly, we used the package $\mathtt{reticulate}$ to import all evaluation routines from the Python scripts provided by the ViEWS team to R \citep{reticulate}.  

\printbibliography